\documentclass[12pt]{article}
\usepackage{jheppub}
\usepackage{mathtools,amssymb,amsthm}
\usepackage[utf8]{inputenc}
\usepackage{enumitem}
\usepackage{xcolor}
\usepackage{graphicx}
\usepackage{multirow}
\usepackage[normalem]{ulem}
\usepackage{longtable}
\allowdisplaybreaks

\newcommand{\be}{\begin{equation}}
\newcommand{\ee}{\end{equation}}
\newcommand{\bi}{\begin{itemize}}
\def \bea {\begin{eqnarray}}
\def \eea {\end{eqnarray}}

\def\ba#1\ea{\begin{align}#1\end{align}}
\def\bad#1\ead{\begin{aligned}#1\end{aligned}}
\def\bg#1\eg{\begin{gather}#1\end{gather}}
\def\bm#1\em{\begin{multline}#1\end{multline}}
\def\bmd#1\emd{\begin{multlined}#1\end{multlined}}

\newcommand{\ignore}[1]{}

\def\a{\alpha}
\def\b{\beta}
\def\c{\chi}

\def\d{\delta}

\def\e{\epsilon}

\def\g{\gamma}
\def\G{\Gamma}

\def\l{\lambda}
\def\L{\Lambda}
\def\m{\mu}
\def\n{\nu}

\def\r{\rho}
\def\s{\sigma}

\def\t{\tau}

\newcommand{\la}{\label}

\newcommand{\re}{\ref}
\newcommand{\er}{\eqref}

\newcommand{\fr}{\frac}

\newcommand{\pa}{\partial}

\newcommand{\wtd}{\widetilde}
\newcommand{\ph}{\phantom}
\newcommand{\eq}{\equiv}

\newcommand{\cd}{\cdots}
\newcommand{\nn}{\nonumber}
\newcommand{\qu}{\quad}
\newcommand{\qqu}{\qquad}
\newcommand{\lt}{\left}
\newcommand{\rt}{\right}
\newcommand{\lra}{\leftrightarrow}

\renewcommand{\(}{\left(}
\renewcommand{\)}{\right)}
\renewcommand{\[}{\left[}
\renewcommand{\]}{\right]}
\newcommand{\<}{\langle}
\renewcommand{\>}{\rangle}

\newcommand{\cO}{{\mathcal O}}

\newcommand{\T}[3]{{#1^{#2}_{\ph{#2}#3}}}
\newcommand{\Tu}[3]{{#1_{#2}^{\ph{#2}#3}}}

\begin{document}

\title{A Tale of Two Butterflies: An Exact Equivalence in Higher-Derivative Gravity}
\author{Xi Dong,}
\author{Diandian Wang,}
\author{Wayne W. Weng,}
\author{and Chih-Hung Wu}
\affiliation{Department of Physics, University of California, Santa Barbara, CA 93106, USA}
\emailAdd{xidong@ucsb.edu}
\emailAdd{diandian@physics.ucsb.edu}
\emailAdd{wweng@physics.ucsb.edu}
\emailAdd{chih-hungwu@physics.ucsb.edu}

\abstract{We prove the equivalence of two holographic computations of the butterfly velocity in higher-derivative theories with Lagrangian built from arbitrary contractions of curvature tensors. The butterfly velocity characterizes the speed at which local perturbations grow in chaotic many-body systems and can be extracted from the out-of-time-order correlator. This leads to a holographic computation in which the butterfly velocity is determined from a localized shockwave on the horizon of a dual black hole. A second holographic computation uses entanglement wedge reconstruction to define a notion of operator size and determines the butterfly velocity from certain extremal surfaces. By direct computation, we show that these two butterfly velocities match precisely in the aforementioned class of gravitational theories. We also present evidence showing that this equivalence holds in all gravitational theories. Along the way, we prove a number of general results on shockwave spacetimes.
}

\maketitle


\section{Introduction}

Chaos is prevalent in the physical world: small changes in initial conditions can lead to drastic variations in the outcome. This sensitive dependence on the initial state is known as the butterfly effect. In classical systems, chaotic dynamics is characterized by an exponential deviation between trajectories in phase space; the Lyapunov exponent parameterizes the degree of deviation.

In quantum mechanical systems, defining chaos is a more challenging task as the wavefunction is governed by a linear evolution~\cite{Berry:1977zz}. Nevertheless, one can characterize quantum chaos by the strength of the commutator $[V(t),W(0)]$ between two generic operators with time separation $t$ \cite{1969JETP...28.1200L,Almheiri:2013hfa}.\footnote{Random matrix theory provides a different way of characterizing quantum chaos from the spectral statistics~\cite{Bohigas:1984}.} One useful measure of the typical matrix elements of this commutator is the expectation value of $\lt|[V(t),W(0)]\rt|^2$ (where the square is defined as $|\cO|^2\eq \cO^\dagger \cO$ for any operator $\cO$).  In a chaotic system, this quantity grows with $t$ and becomes significant around the scrambling time $t_*$~\cite{Hayden:2007cs,Sekino:2008he}, behaving like $e^{\lambda_L (t-t_*)}$. Here $\lambda_L$ is the (quantum) Lyapunov exponent. 

In this paper, we are interested in the spreading of the quantum butterfly effect in spatial directions.  We therefore specialize $W$ and $V$ into local operators with spatial separation $x$ in $d-1$ spatial dimensions.  The corresponding expectation value behaves like~\cite{Shenker:2013pqa,Roberts:2014isa,Roberts:2016wdl}
\be
C_\beta(x,t) \equiv \langle \lt| [V(x,t),W(0,0)] \rt|^2 \rangle_\beta \sim e^{\lambda_L (t-t_* - |x|/v_B)},
\ee
where $\langle \, \cdot \, \rangle_\beta$ denotes the expectation value in a thermal state with inverse temperature $\beta$.  The quantity $C_\beta(x,t)$ characterizes the strength of the butterfly effect at $(x,t)$ detected by a probe $V$ following an earlier perturbation $W(0,0)$.

Here $v_B$ is known as the {\it butterfly velocity}.  It is the speed at which the region with $C_\beta(x,t) \gtrsim 1$ expands outward.  Intuitively, the commutator probes how a small perturbation $W(0,0)$ spreads over the system. For two operators that are far separated, the commutator is zero at early times and becomes large at sufficiently late times. In particular, the region in which $C_\beta(x,t)$ is order unity gives a measure of the size of the operator $W(0,0)$ at time $t$, and the speed at which the size grows over time is precisely the butterfly velocity. 

In recent years, there has been much interest in the role of chaotic dynamics in holographic quantum systems, particularly in the context of AdS/CFT~\cite{Shenker:2013pqa, Shenker:2013yza, Roberts:2014isa, Shenker:2014cwa, Roberts:2016wdl, Perlmutter:2016pkf, Jensen:2016pah}. In this context, a thermal state in the boundary CFT can be realized as a black hole in the bulk AdS spacetime~\cite{Maldacena:2001kr}. The rapid thermalization of a local perturbation on the boundary can be understood from the fast scrambling dynamics of the black hole~\cite{Hayden:2007cs,Sekino:2008he,Susskind:2011ap,Lashkari:2011yi}. In particular, the Lyapunov exponent is related to the exponential blueshift of early infalling quanta in the near-horizon region. For CFTs dual to Einstein gravity (possibly corrected by a finite number of higher-derivative terms), the Lyapunov exponent is universal and saturates~\cite{Shenker:2013pqa, Shenker:2013yza, Roberts:2014isa, Shenker:2014cwa} the chaos bound $\lambda_L \leq 2\pi/\beta$~\cite{Maldacena:2015waa}. This can be understood from the universality of the near-horizon Rindler geometry. If one considers perturbations sent from sufficiently far in the past, the quanta are significantly blue-shifted by the Rindler geometry. At around a scrambling time, the backreaction on the background geometry can be described by a shockwave on the horizon~\cite{Shenker:2013pqa, Shenker:2013yza, Shenker:2014cwa}. The strength of the shockwave grows exponentially as we insert the perturbation at earlier and earlier times. 

This observation leads to one way of obtaining the butterfly velocity. To see it concretely, consider a localized perturbation in the thermofield-double (TFD) state dual to a two-sided $(d+1)$-dimensional planar black hole. Inserting such a spatially localized perturbation on one boundary corresponds to injecting a small number of quanta which then proceed to fall towards the black hole in the bulk. The result of doing so is a localized shockwave~\cite{Roberts:2014isa}. The spatial region in which the shockwave has non-trivial support defines a size of the corresponding boundary operator. As we send the perturbation at earlier and earlier times, the size of this region grows, and this ``speed of propagation" determines a butterfly velocity $v_B$. 

An alternative way of calculating the operator size and the butterfly velocity in holography was proposed in~\cite{Mezei:2016wfz}.  It is based on entanglement wedge reconstruction, which states that a bulk operator within the entanglement wedge of any boundary subregion can be represented by some boundary operator on that subregion~\cite{Czech:2012bh,Wall:2012uf,Headrick:2014cta,Dong:2016eik}. Consider again a local operator inserted on the boundary, which in the bulk can be thought of as a particle falling into the black hole. By entanglement wedge reconstruction, a boundary region whose entanglement wedge contains the particle possesses full information about the boundary operator. The smallest spherical region that does so defines a notion of size for the boundary operator. At early times, the particle is near the boundary; the boundary region, and hence the operator size, is small. At very late times, say after a scrambling time, the particle is very close to the horizon and the entanglement wedge needs to extend deep into the bulk. The corresponding boundary region, and hence the operator size, is very large and in fact grows linearly with time. The corresponding speed, which we denote as $\wtd v_B$, quantifies the growth of the operator size.  It provides a second way of computing the butterfly velocity holographically.

These two holographic computations of the butterfly velocity appear to be very different and unrelated to each other.  However, it was shown directly in~\cite{Mezei:2016wfz} that the results of the two computations agree for Einstein gravity and for higher-derivative gravity with up to four derivatives on the metric.

The goal of this paper is to prove that the two computations of the butterfly velocity continue to agree in general higher-derivative theories of gravity. We will focus on the family of general $f(\text{Riemann})$ theories, namely those with Lagrangians built from arbitrary contractions of an arbitrary number of Riemann tensors:\footnote{Nevertheless, in Section~\re{sec:discussion} we will discuss one example that is more general than $f(\text{Riemann})$ theories, where the two butterfly velocities continue to agree.}
\be\label{eq:HDT}
\mathcal{L} = \frac{1}{2}(R -2\L) + \lambda_1 R^2 + \lambda_2 R_{\mu\nu}R^{\mu\nu} + \lambda_3 R_{\mu\nu\rho\sigma}R^{\mu\nu\rho\sigma} + \lambda_4 R^3 + \cdots,
\ee
where the higher-derivative terms are viewed as perturbative corrections to the leading Einstein-Hilbert action. The main purposes of studying higher-derivative theories are twofold: (1) they arise generally as perturbative corrections to Einstein gravity in low energy effective theories of UV-complete models of quantum gravity such as string theory; (2) the agreement between two computations of the butterfly velocity for general higher-derivative theories would suggest an equivalence between the two methods themselves rather than a coincidence in certain theories. This equivalence then suggests a deeper connection between gravitational shockwaves and holographic entanglement, as well as providing further evidence for entanglement wedge reconstruction.

The rest of the paper is organized as follows. In Section~\ref{sec:methods}, we begin with a detailed review of the two holographic calculations of the butterfly velocity. In Section~\ref{sec:butterfly}, we derive general expressions for the two butterfly velocities in $f($Riemann$)$ theories. In Section~\ref{sec:proof}, we prove $v_B = \wtd v_B$ for this class of theories, which is our main result. In Section~\ref{sec:discussion}, we end with a discussion of this result and comment on potential future directions.

\section{Operator size and the butterfly velocity}
\label{sec:methods}
In chaotic many-body systems, the size of generic local operators --- the spatial region on which the operator has large support --- grows ballistically under Heisenberg time evolution.  More specifically, consider a perturbation by such a local operator.  Under chaotic evolution, information about the perturbation is {\it scrambled} amongst the local degrees of freedom and spreads throughout the system. The maximum speed at which this occurs is the butterfly velocity $v_B$. It can be regarded as a finite temperature analogue of the Lieb-Robinson bound~\cite{Lieb:1972wy,Roberts:2016wdl}. The existence of this bound on information scrambling is a general feature of chaotic systems.

One can define operator size more precisely using the square of the commutator for two generic local operators $W$ and $V$:
\begin{equation}\label{eq:commutator}
\begin{split}
    C_\beta(x,t_W) \equiv &\, \langle  [V(x,0),W(0,-t_W)]^\dagger [V(x,0),W(0,-t_W)] \rangle_\beta\\
=&\, 2-2\,\text{Re}\, \langle V(x,0)^\dagger W(0,-t_W)^\dagger V(x,0) W(0,-t_W) \rangle_\beta,
\end{split}
\end{equation}
where $t_W>0$ so that $W$ is inserted at an earlier time than $V$ and the expectation value is taken in a thermal state with inverse temperature $\beta$.\footnote{In going to the second line of~\eqref{eq:commutator}, we have assumed the two operators $W$ and $V$ to be unitary. This is not a crucial assumption, but does simplify our discussion.} The second term in the second line is called an out-of-time-order correlator (OTOC) and carries all of the non-trivial information in \eqref{eq:commutator}. The exponential decay of the OTOC over time is commonly used as an indicator of chaos in quantum many-body systems.

\begin{figure}[t]
    \centering
    \includegraphics[height=1.8in]{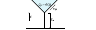}
    \caption{Illustration of the butterfly cone. A localized perturbation at some initial time scrambles everything at that location after a scrambling time $t_*$. This effect then spreads out spatially at the butterfly velocity $v_B$, defining the butterfly cone where $C_\b$ is of order one.}
    \label{fig:cone2}
\end{figure}

Under chaotic time evolution, the commutator with $x=0$ exhibits an exponential growth near the scrambling time $t_*$,\footnote{For example, in maximally chaotic many-body systems with $\mathcal{O}(N)$ degrees of freedom per site, $t_*$ is approximately $\frac{\beta}{2\pi}\log N$.} defined as the time at which the commutator becomes of order unity. Now considering non-zero $x$, information from a localized perturbation is scrambled among the local degrees of freedom and spreads throughout the system at the butterfly velocity $v_B$. In this scrambling regime, the behavior of the commutator has the following universal form \cite{Roberts:2016wdl},
\begin{equation}
    C_\beta(x, t_W) \sim e^{\lambda_{L}\left(t_W-t_*-|x| / v_{B}\right)}.
\end{equation}
The Lyapunov exponent $\lambda_L$ gives a time scale for scrambling at a fixed spatial location, while $v_B$ parametrizes the delay in scrambling due to spatial separation (see Figure~\ref{fig:cone2}). At time $t_W$ after the insertion of the $W$ perturbation, the commutator is order unity in the spatial region defined by $|x| \leq v_B(t_W-t_*)$, and is exponentially suppressed outside this region. This gives a precise notion of the size of an operator under time evolution. 

A nice way to think about the OTOC, and hence the commutator, is as the overlap $\<\psi'|\psi\>$ between the two states
\be
\label{eq:psi_psiprime}
|\psi\> = V(x,0)W(0,-t_W)|\text{TFD}\>_\b \qu \text{and} \qu |\psi'\> = W(0,-t_W)V(x,0)|\text{TFD}\>_\b.
\ee
Here $|\text{TFD}\>_\b$ is the thermofield-double state:
\be
|\text{TFD}\>_\b \eq \fr{1}{Z(\b)^{1/2}} \sum_n e^{-\b E_n/2} |n\>_L |n\>_R,
\ee
where $L$ denotes the original system and $R$ denotes an identical copy, $|n\>$ is a complete set of energy eigenstates with energy $E_n$, and $Z(\b)$ is the thermal partition function at inverse temperature $\b$. The states \er{eq:psi_psiprime} are obtained from $|\text{TFD}\>_\b$ by acting with operators in the $L$ system. In particular, the state $|\psi\>$ corresponds to acting with $W$ in the past at $t=-t_W$ and time evolving to $t=0$ before inserting an operator $V$. The state $|\psi'\>$ corresponds to creating a perturbation $V$ at $t=0$, time evolving to the past, inserting $W$, and finally evolving back to $t=0$. Under chaotic dynamics, the operator $W$ is scrambled amongst degrees of freedom in its neighborhood. If $W$ is inserted far enough in the past, it can interfere with the perturbation due to $V$ and prevent it from reappearing at $t=0$ in the state $|\psi'\>$. Consequently, the state $|\psi'\>$ would have a small overlap with $|\psi\>$ since the latter has $V$ inserted at $t=0$ by construction. This small overlap means that the commutator \er{eq:commutator} is of order one, which defines the butterfly cone.

We now review two methods of calculating the butterfly velocity in holographic systems, which we call the shockwave method and the entanglement wedge method. 

\subsection{The shockwave method}
\label{ssec:rev_shock}
\begin{figure}[t]
    \centering
    \includegraphics[width=0.8\textwidth]{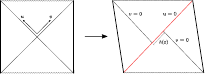}
    \caption{An illustration of the shockwave spacetime. By sending in a localized energy packet from the left boundary around a scrambling time in the past, the backreaction is described by a localized shockwave on the horizon at $u = 0$ with profile $h(x)$.}
    \label{fig:shockwave}
\end{figure}

One way of capturing the chaotic behavior in a holographic CFT is to introduce a perturbation at the asymptotic boundary and study the backreaction to the geometry as it propagates into the bulk. We will consider two copies of the CFT in a thermofield double state, which is dual to a two-sided eternal black hole in the bulk. Now let us act on the left boundary CFT with a local operator $W$ at the origin $x=0$ and boundary time $-t_W$:
\be
W(0,-t_W) |\text{TFD} \rangle= e^{-iH_L t_W}W(0,0) e^{iH_L t_W}| \text{TFD} \rangle.
\ee
In the bulk, this perturbation corresponds to inserting an energy packet near the asymptotic boundary, which then falls towards the black hole. If we take $t_W$ to be large, by the time it reaches $t=0$, it will have gained considerable energy due to the exponentially large blueshift near the horizon, which then backreacts significantly on the spacetime. For large enough $t_W$,\footnote{For the perturbation to be large (but not Planckian), the time at which we send in the particle needs to be around the scrambling time, $|t_W| \approx t_\ast$ \cite{Hayden:2007cs, Sekino:2008he, Susskind:2011ap, Lashkari:2011yi, Shenker:2013pqa}.} the backreacted geometry is well-described by a shockwave along the horizon as shown in Figure~\ref{fig:shockwave}.

To be more specific, let us consider a general $(d+1)$-dimensional planar black hole. The metric can be written in Kruskal coordinates as
\be \label{eq:bg_metric_kruskal}
ds^2=2A(uv) dudv+B(uv) dx^i dx^i,
\ee
where the functions $A(uv)$ and $B(uv)$ in general depend on the gravitational theory and the matter content, and $i\in\{1,\dots,d-1\}$ labels the transverse directions. The two horizons live at $u = 0$ and $v = 0$. We will also rescale the transverse directions so that $B(0)=1$. 

A localized shockwave on the horizon at $u=0$ is sourced by a change in the stress tensor due to a perturbation with initial asymptotic energy $\mathcal{E}$
\begin{align}
\label{eq:shock_stress}
    \d T^v_u&=A^{-1}\mathcal{E} e^{\frac{2 \pi}{\beta} t_W} \d(u) \d(x).
\end{align}
The prefactor $\mathcal{E}e^{\frac{2 \pi}{\beta} t_W} $ can be thought of as the effective energy of the blueshifted perturbation. Note that $t_W$ is not a spacetime coordinate but rather parameterizes the time at which the perturbation is inserted. Inserting the perturbation at earlier times increases this effective energy and results in a larger backreaction. 

To solve for the backreaction, it is sufficient to perturb only the $uu$ component of metric by an amount parameterized by some function $h(x)$, 
\be \label{eq:pl}
ds^2=2A(uv) dudv+B(uv) dx^i dx^i-2A(uv)h(x) \d(u) du^2.
\ee
The function $h(x)$ --- which we will refer to as the \textit{shockwave profile} --- is determined by the equations of motion. Assuming that the equations of motion for the unperturbed background solution \eqref{eq:bg_metric_kruskal},
\begin{equation}
    E_{\m\n} \( \equiv \frac{2}{\sqrt{-g}} \frac{\d S}{\d g^{\m\n}} \) = T_{\m\n},
\end{equation}
are satisfied (where $S$ is the gravitational part of the action), it suffices to consider the perturbed equations of motion
\begin{equation}\label{eq:pert_EOM}
    \d E_{\mu}^{\nu} = \delta T_{\mu}^{\nu}
\end{equation}
sourced by the shockwave stress tensor \eqref{eq:shock_stress}. For Einstein gravity, this shockwave equation of motion was first derived for a vacuum background by Dray and t'Hooft \cite{Dray:1984ha}, and later generalized to allow a non-trivial background stress-tensor in \cite{Sfetsos:1994xa}.

Now we consider general theories of gravity.  As we prove in Appendix~\ref{app:lin}, the only non-trivial component of \eqref{eq:pert_EOM} is
\begin{align}\label{eq:shock_EOM_2}
    \delta E^{v}_u =\d T^{v}_{u} = A^{-1}\mathcal{E}e^{\frac{2 \pi}{\beta} t_W} \delta(u) \delta(x).
\end{align}
Furthermore, in the same appendix we show that \eqref{eq:shock_EOM_2} truncates automatically at linear order in $h(x)$ (and its $x^i$-derivatives). Indeed, the equations of motion reduce to a single ODE for the shockwave profile $h(x)$, which we will refer to as the \textit{shockwave equation}. As we will show, \eqref{eq:shock_EOM_2} can be solved for large $r=|x|$ with the following ansatz for the shockwave profile
\be
\label{eq:h_ansatz}
h(x)\sim\frac{e^{\frac{2 \pi}{\beta}t_W-\mu r}}{r^{\#}},
\ee
where $\mu > 0$ and $\#$ is some integer that will not be important.

\begin{figure}[t]
    \centering
    \includegraphics[width=0.8\textwidth]{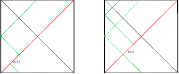}
    \caption{Bulk representations of $|\psi\rangle$ and $|\psi^\prime\rangle$. For the state $|\psi\rangle$ (left), the shockwave (red) created by $W$ is inserted first and the probe operator $V$ is then inserted at $t=0$, manifesting itself as a particle (green) propagating into the future and the past. On its way to the past, it encounters the already existent shockwave and is shifted by an amount $h(x)$. On the other hand, $|\psi^\prime\rangle$ (right) is created by inserting $V$ first and then placing the shockwave $W$. The shockwave alters the original trajectory of the $V$ particle, such that it is shifted to the future of the shockwave.}
    \label{fig:psi}
\end{figure}

To see the connection to the OTOC, and hence the butterfly velocity, let us consider the overlap of the two states defined in \eqref{eq:psi_psiprime} from the bulk perspective (see Figure~\ref{fig:psi}). The states differ in the trajectory of the particle $V$, due to the shift $h(x)$ from the shockwave $W$ occurring either in the past or future evolution. Because the only difference between the two states is in the particle trajectory, the size of the overlap is controlled by the shift: it is close to unity when the $h$ is small and close to zero when the $h$ is large. Let us for concreteness define the boundary of the butterfly cone to be where $C_\beta= 1$, which translates to $\text{Re}\,\langle \psi^\prime|\psi\rangle= \frac{1}{2}$. This corresponds to some threshold value of the shift, say $h(x)=h_*$. According to \eqref{eq:h_ansatz}, the size of the butterfly cone then grows as a function of $t_W$ with the velocity
\be\label{eq:vB}
v_B=\frac{2\pi}{\beta \mu}
\ee
at large $r$. We identify this with the butterfly velocity. Solving the equation of motion \eqref{eq:shock_EOM_2} yields a value for $\mu$ and thus a value for $v_B$ in terms of the functions $A(uv)$ and $B(uv)$ in the background metric \eqref{eq:bg_metric_kruskal}. \\

Let us now look at some examples. 

\paragraph{Einstein gravity} Let us start with the simplest case of Einstein gravity, $\mathcal{L}_{\text{EH}}=\frac{1}{2}(R-2\Lambda)$. The calculations follow that in \cite{Shenker:2013pqa,Roberts:2014isa}, which we now review. Using \er{eq:shock_EOM_2}, the shockwave equation is given by
\begin{equation}\la{eq:seomeh}
    \delta E^v_u=\frac{1}{B}\(\partial_i \partial_i  -\fr{d-1}{2} \frac{B'}{A}  \) h(x)\d(u) =A^{-1} \mathcal{E} e^{\frac{2 \pi}{\beta} t_W} \d(x)\d(u).
\end{equation}
In the large-$r$ limit, we can then substitute the ansatz \er{eq:h_ansatz}, and obtain an algebraic equation for the parameter $\mu$,
\begin{align}\la{eq:einmu}
\mu^2  -\fr{d-1}{2} \frac{B_1}{A_0}=0,
\end{align}
where $A_0\equiv A(0)$ and $B_1\equiv B^\prime(0)$. Choosing the positive root for $\mu$ and using \eqref{eq:vB}, we therefore find the butterfly velocity
\begin{equation}\label{eq:vB_ein}
    v_B=\frac{2\pi}{\beta}\sqrt{\frac{2A_0}{(d-1)B_1}}.
\end{equation}

Note that the equation for $\m$ and thus the butterfly velocity depend only on the behavior of the metric near the $u = 0$ horizon. This is enforced by the overall factor of $\d(u)$ in the equation of motion. As we will see, in the entanglement wedge method, this feature will be reproduced via a different mechanism --- by taking a near-horizon limit of an extremal surface. This is one of the many distinctions between the two methods, making their agreement quite non-trivial.

\paragraph{Lovelock gravity} As was shown in \cite{Roberts:2014isa}, for Gauss-Bonnet gravity whose Lagrangian is the Einstein-Hilbert term $\mathcal{L}_{\text{EH}}$ plus
\be
\mathcal{L}_{\text{GB}}=\lambda_{\text{GB}}(R^2-4 R_{\mu \nu} R^{\mu \nu}+R_{\mu \nu \rho \sigma} R^{\mu \nu \rho \sigma}),
\ee
the coupling constant $\lambda_{\text{GB}}$ does not contribute to the shockwave equation \eqref{eq:seomeh}. 

We can further show that this is the case for Lovelock gravity, a general $2p$-derivative theory with second order equations of motion. The Lagrangian is $\mathcal{L}_{\text{EH}}$ plus
\be\label{eq:lovelock}
\mathcal{L}_{2p} = \frac{\lambda_{2p}}{2^p}\d^{\mu_1\nu_1\cd \mu_p\nu_p}_{\rho_1\sigma_1\cd \rho_p\sigma_p} \Tu R{\mu_1\nu_1}{\rho_1\sigma_1} \cd \Tu R{\mu_p\nu_p}{\rho_p\sigma_p},
\ee
where the generalized delta symbol is a totally antisymmetric product of Kronecker deltas, defined recursively as
\begin{equation}
\delta_{\nu_{1} \nu_{2} \cdots \nu_{n}}^{\mu_{1} \mu_{2} \cdots \mu_{n}}=\sum_{i=1}^{n}(-1)^{i+1} \delta_{\nu_{i}}^{\mu_{1}} \delta_{\nu_{1} \cdots \hat{\nu}_{i} \cdots \nu_{n}}^{\mu_{2} \mu_{3} \cdots \mu_{n}}.
\end{equation}
Choosing $p=1$ gives Einstein-Hilbert, while $p=2$ reduces to Gauss-Bonnet. The metric equation of motion is given by
\be
E^\alpha_\beta = -\frac{\lambda_{2p}}{2^{p}}\d^{\alpha \mu_1\nu_1\cd \mu_p\nu_p}_{\beta \rho_1\sigma_1\cd \rho_p\sigma_p} \Tu R{\mu_1\nu_1}{\rho_1\sigma_1} \cd \Tu R{\mu_p\nu_p}{\rho_p\sigma_p}.
\ee
We are interested in $\delta E^v_u$.  It is not difficult to show that the only way to get a non-zero contraction is to make one of the Riemann tensors $\Tu R{ui}{vj}$ and the rest $\Tu R{kl}{mn}$. Using our metric ansatz \er{eq:pl}, one can show that $\delta\Tu R{kl}{mn}=0$, so any potential contribution can only come from $\delta \Tu R{ui}{vj} \propto \d(u)$ multiplied by $p-1$ factors of $\Tu R{kl}{mn}$ (see Eq.~\eqref{eqs:riemanns} for the detailed expressions). The latter vanishes on the horizon of the planar black hole, due to the flatness of the transverse directions. Hence, we find that as long as $p>1$ the Lovelock corrections do not contribute to the shockwave equation \eqref{eq:seomeh}.

\subsection{The entanglement wedge method} \label{ssec:EW}

In holography, the butterfly effect manifests itself in another way as first observed in \cite{Mezei:2016wfz}. The intuition comes from entanglement wedge reconstruction \cite{Czech:2012bh,Wall:2012uf,Headrick:2014cta,Dong:2016eik}, which states that a given boundary region contains all of the information inside its entanglement wedge, which is a bulk region bounded by the boundary region and the corresponding extremal surface. As was argued in \cite{Mezei:2016wfz}, this allows us to define a notion of operator size on the boundary.

Consider a thermal state in the boundary CFT dual to a planar black hole in the bulk. Let us perturb the boundary state by acting with a local operator. Under the chaotic time evolution, information from the perturbation is scrambled throughout the system. At late times, the boundary region over which this information is smeared propagates outwards at a constant velocity. In the bulk, the perturbation corresponds to a probe particle (or wavepacket) originating from the asymptotic boundary and falling towards the black hole, its trajectory determined from our choice of the bulk theory and the background geometry. According to entanglement wedge reconstruction, any boundary region whose entanglement wedge contains the particle should contain all the information of the corresponding boundary operator. In particular, we would like to consider the extremal surface which barely encloses the particle in its entanglement wedge. The corresponding boundary region then defines a size for the boundary operator.

\begin{figure}[t]
    \centering
    \includegraphics[width=1.00\textwidth]{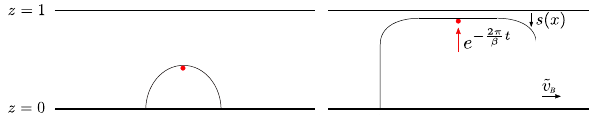}
    \caption{An illustration of the entanglement wedge method. A particle is released from the asymptotic boundary and we show the RT surface which barely encloses the particle. At early times (left), the profile of the RT surface depends sensitively on the metric away from the horizon; but at late times (right), the RT surface in the near-horizon region has a characteristic profile which propagates outwards at the butterfly velocity $\wtd v_B$.}
    \label{fig:EW}
\end{figure}

To extract the butterfly velocity, we now study how this extremal surface changes as it follows the trajectory of the particle. Note that even though the location of the particle is time-dependent, the background spacetime is static\footnote{Here we do not need to analyze the backreaction of the particle, unlike in the shockwave method.} and at any given time we may use a Ryu-Takayanagi (RT) surface \cite{Ryu:2006bv,Ryu:2006ef} (instead of its dynamical generalization --- the HRT surface \cite{Hubeny:2007xt}). At early times, the shape of the RT surface will depend sensitively on details of the background metric. However, at late times, the surface approaches the near-horizon region and exhibits a characteristic profile which propagates outwards at a constant velocity (see Figure \ref{fig:EW}). This velocity, which we can identify as the butterfly velocity $\wtd v_B$, depends only on the bulk theory and the near-horizon geometry of the black hole (as long as the theory admits a black hole solution, which we can ensure by taking the coefficients of higher-derivative terms to be small so that a solution perturbatively close to the static planar black hole in Einstein gravity exists).

In general higher-derivative gravity, the location of the RT surface is determined by extremizing the holographic entanglement entropy functional $S_{\text{EE}}$ among all bulk surfaces homologous to the corresponding boundary region \cite{Dong:2013qoa,Camps:2013zua,Dong:2017xht,Dong:2019piw}.  For the $f(\text{Riemann})$ theories that we focus on, the entropy functional $S_{\text{EE}}$ can be found in \cite{Dong:2013qoa}.  The important terms for our purpose is
\begin{multline}
\label{eq:S_EE}
S_{\text{EE}}
=2 \pi \int d^{d-1} y \sqrt{\gamma}\bigg\{-\frac{\partial \mathcal{L}}{\partial R_{\mu \rho \nu \sigma}} \varepsilon_{\mu \rho} \varepsilon_{\nu \sigma}- \frac{\partial^{2} \mathcal{L}}{\partial R_{\mu_{1} \rho_{1} \nu_{1} \sigma_{1}} \partial R_{\mu_{2} \rho_{2} \nu_{2} \sigma_{2}}} 2 K_{\lambda_{1} \rho_{1} \sigma_{1}} K_{\lambda_{2} \rho_{2} \sigma_{2}} \times \\
\times\left[\left(n_{\mu_{1} \mu_{2}} n_{\nu_{1} \nu_{2}}+\varepsilon_{\mu_{1} \mu_{2}} \varepsilon_{\nu_{1} \nu_{2}}\right) n^{\lambda_{1} \lambda_{2}}-\left(n_{\mu_{1} \mu_{2}} \varepsilon_{\nu_{1} \nu_{2}}+\varepsilon_{\mu_{1} \mu_{2}} n_{\nu_{1} \nu_{2}}\right) \varepsilon^{\lambda_{1} \lambda_{2}}\right] + \cd\bigg\},
\end{multline}
where $y$ denotes a set of coordinates on an appropriate codimension-2 surface, $\gamma$ is the determinant of its induced metric, $K_{\l\r\s}$ is its extrinsic curvature tensor, $n_{\m\n}$ is the induced metric (and $\varepsilon_{\m\n}$ is the Levi-Civita tensor) in the two orthogonal directions while vanishing in the remaining directions, and $\cd$ denotes terms that are higher-order\footnote{These higher-order terms are difficult to write down explicitly because of `splitting' \cite{Dong:2013qoa,Miao:2014nxa,Miao:2015iba,Camps:2016gfs}, although they can in principle be determined by using appropriate equations of motion \cite{Dong:2013qoa,Dong:2017xht,Dong:2019piw}.  Fortunately, here we only need $S_{\text{EE}}$ up to second order in $K$ (and its derivatives), which can be obtained by setting $q_\a=0$ in Eq.~(3.30) of \cite{Dong:2013qoa} and is free from the splitting difficulty --- this follows roughly from the results of \cite{Dong:2013qoa} but will also be proved carefully in a separate work \cite{upcoming}.  Moreover, the same $S_{\text{EE}}$ up to second order in $K$ (and its derivatives) was derived using the second law of black hole thermodynamics \cite{Wall:2015raa}.} in $K_{\l\r\s}$ and its derivatives than the second-order term shown here.  As we will explain in Section~\re{sec:butterflyew}, these higher-order terms do not affect the calculation of the butterfly velocity.  Note that Eq.~\er{eq:S_EE} works in Lorentzian signature, which we obtain by analytically continuing the corresponding Euclidean expression via $\mathcal{L}\to -\mathcal{L}, n_{\mu\nu}\to n_{\mu\nu}, \varepsilon_{\mu\nu}\to -i\varepsilon_{\mu\nu}$.\footnote{The extra factor of $-i$ can be traced back to the definition $\varepsilon_{\mu\nu}=\varepsilon_{ab}n^{(a)}_\mu n^{(b)}_\nu$ where $\varepsilon_{ab}=\sqrt{g}\, \wtd{\varepsilon}_{ab}$, with $\wtd{\varepsilon}_{ab}$ being the Levi-Civita symbol. Going to the Lorentzian version which comes with $\sqrt{-g}$ requires a factor of $i$. The minus sign is a result of going from the convention in \cite{Dong:2013qoa} where $\wtd\varepsilon_{\tau x}=-1$ to the standard Lorentzian convention $\wtd\varepsilon_{tx}=1$.}  Extremizing $S_{\text{EE}}$ leads to a differential equation for the location of the RT surface, which we refer to as the \textit{RT equation}.

For a general planar black hole, the metric is given in \er{eq:bg_metric_kruskal}, which can be written in Poincar\'e-like coordinates as
\be\label{eq:bg_metric_poincare}
d s^{2}=\frac{1}{z^{2}}\left[-\frac{f(z)}{h(z)} d t^{2}+\frac{d z^{2}}{f(z)}+dx^{i}dx^i\right],
\ee
where the horizon is located at $z=1$. In the near-horizon region, we can Taylor expand the functions $f(z) = f_1(1-z) + f_2(1-z)^2 + \cd$ and $h(z) = h_0 + h_1(1-z) + h_2(1-z)^2 + \cd$. As in \cite{Mezei:2016wfz}, we will consider spherical boundary regions, for which the corresponding RT surfaces are also spherical. We use $r=|x|$ to denote the radial coordinate in the $x^i$ directions. This reduces the RT equation to a simple ODE.

At late times, the probe particle is exponentially close to the horizon and follows a trajectory $1-z(t) \sim e^{-\frac{4\pi}{\beta} t}$ given by the geodesic equation. As such, we will focus on the near-horizon region near $z = 1$. In particular, defining the RT surface by $z = Z(x)$, we can parametrize the near-horizon limit by writing
\be
\label{eq:Z_ito_S}
Z(x) = 1 - \e s(x)^2,
\ee
where $\e$ is a small positive number and we have defined a new function $s(x)$ which we call the \textit{RT profile}. The profile is determined by solving the RT equation, which we Taylor expand around $\e = 0$. The RT surface will stay close to the horizon for small $r$ and start to depart the near-horizon region at some large radius $r=r_*$, where $\e s(x)^2|_{r=r_*} \sim \mathcal{O}(1)$, after which it reaches the asymptotic boundary within an order one distance. Thus, the corresponding boundary region has approximately radius $r_*$ and this defines a size of the boundary operator. To model this behavior, one can use an ansatz\footnote{This ansatz is valid when $r$ is large enough (so that we may ignore higher-order corrections in $1/r$) but not too large (so that the RT surface has not exited the near-horizon region). This regime of validity is parametrically large for a large boundary region.}
\be
\label{eq:S_ansatz}
s(x) \sim \frac{e^{\wtd\m r}}{r^\#},
\ee
where $\wtd \m > 0$ and $\#$ is some unimportant integer.

At each point in time, we demand the tip of the RT surface intersects the particle, which is enforced by setting $s(x=0,t)\sim e^{-\frac{2\pi}{\beta}t}$. Therefore, the time-dependent RT profile is given by
\be
s(x,t) \sim \frac{e^{\wtd\m r-\frac{2\pi}{\b}t}}{r^\#}.
\ee
At any given time $t$, there is some radius $r = r_*(t)$ such that $\e s(x,t)^2|_{r=r_*(t)} \sim \mathcal{O}(1)$. This in turn gives the radius of the boundary region (modulo an order one distance), which propagates outward with characteristic velocity
\be\label{eq:vBtd}
\wtd{v}_B\equiv \frac{2\pi}{\b \wtd\m}.
\ee
Solving the RT equation yields a value for $\wtd \m$ and thus a value for $\wtd v_B$ in terms of the functions $f(z)$ and $h(z)$ in the background metric \eqref{eq:bg_metric_poincare}. This is the second way of calculating the butterfly velocity. \\

Let us now revisit the examples in Section~\ref{ssec:rev_shock} using the entanglement wedge method.

\paragraph{Einstein gravity}
For Einstein gravity with $\mathcal{L}_\text{EH}=\frac{1}{2}(R-2\Lambda)$, the entropy functional is simply given by the area:
\begin{equation}
    S_\text{EE}=2\pi \int d^{d-1}y \sqrt{\gamma}.
\end{equation}
The entire dependence of this functional on the choice of the surface is then in the determinant of the induced metric $\g$. Using \er{eq:Z_ito_S} and expanding to linear order in $\epsilon$, the induced metric of the surface is given by
\be\label{eq:induced_Zcoord}
\g_{ij}dx^idx^j = \[1+2\epsilon\(s(r)^2 + \fr{2}{f_1}s'(r)^2\)\] dr^2 + r^2\Big(1+2\epsilon s(r)^2\Big) d\Omega_{d-2}^2 + \mathcal{O}(\epsilon^2),
\ee
where we have Taylor expanded $f(z) \equiv f_{1}(1-z)+f_{2}(1-z)^{2}+\cdots$ near the horizon. This is a flat metric up to $\mathcal{O}(\epsilon)$ corrections, which is consistent with the fact that we are expanding near the horizon of a planar black hole. The determinant is then given by
\be \label{eq:inducedDet}
    \sqrt{\gamma} = r^{d-2} + \epsilon\, r^{d-2}\left((d-1)s(r)^{2}+\fr{2}{f_1}s^{\prime}(r)^{2}\right)+\mathcal{O}(\epsilon^2).
\ee
The RT equation is then determined by varying with respect to $s(r)$, which at leading order in $\epsilon$ gives
\begin{equation}
 (d-1)s-\frac{2}{f_1}\left(s^{\prime\prime}+(d-2)\frac{s^\prime}{r} \right)=0.
\end{equation}
Substituting the ansatz \er{eq:S_ansatz} and dropping higher order terms in $1/r$ then gives
\begin{equation}\la{eq:einmutd}
\wtd\mu^{2}=\frac{d-1}{2} f_{1} = \frac{d-1}{2} \frac{B_1}{A_0},
\end{equation}
where the second equality follows from a coordinate transformation. Using \er{eq:vBtd}, we thus find that the resulting butterfly velocity matches the shockwave result \er{eq:vB_ein}.

\paragraph{Lovelock gravity}
We will now show that the Lovelock corrections $\mathcal{L}_{2p}$ with $p>1$ do not contribute to the RT equation, consistent with the shockwave calculation in Section~\ref{ssec:rev_shock}.

It is well-known that the entropy functional for this theory is given by the Jacobson-Myers formula \cite{Jacobson:1993xs}
\be \label{eq:JM}
S_{\text{EE},2p} = 2\pi \lambda_{2p}\int d^{d-1}y \, \sqrt{\gamma} \frac{p}{2^{p-2}}\, \delta^{i_1 j_1\cdots i_{p-1} j_{p-1}}_{k_1 l_1\cdots k_{p-1} l_{p-1}} \mathcal{R}_{i_1 j_1}{}^{k_1 l_1}\cdots \mathcal{R}_{i_{p-1} j_{p-1}}{}^{k_{p-1} l_{p-1}},
\ee
where $\mathcal{R}_{ij}{}^{kl}$ is the intrinsic curvature of the codimension-2 surface. For an RT surface perturbatively close to the horizon, the induced metric is again given by \er{eq:induced_Zcoord}. We can therefore immediately write $\mathcal{R}_{ij}{}^{kl} \sim \mathcal{O}(\epsilon)$, which implies
\be
S_{\text{EE},2p} \sim \mathcal{O}(\epsilon^{p-1}).
\ee
For $p>2$, these terms are higher order in $\epsilon$ and therefore do not contribute. Thus these Lovelock corrections do not modify \er{eq:einmutd}.

For the case of $p = 2$, namely Gauss-Bonnet gravity, a more refined argument is required. In this case, the entropy functional \er{eq:JM} depends on the intrinsic curvature scalar $\mathcal{R} = \mathcal{R}_{ij}{}^{ij}$ which, up to unimportant numerical factors, is given by
\be
\mathcal{R} \sim \fr{\epsilon}{r} \fr{d}{dr} \Big(rs(r)s'(r)\Big) + \mathcal{O}(\epsilon^2),
\ee
where we have neglected terms that only contribute higher orders in $1/r$ to the RT equation compared to the Einstein-Hilbert term \er{eq:inducedDet}. The entropy is then given by
\ba
S_{\text{EE},\text{GB}} &\sim \int d^{d-1}y \sqrt{\gamma} \, \mathcal{R} \sim \epsilon \!\int dr \, r^{d-3} \fr{d}{dr} \Big(rs(r)s'(r)\Big) + \mathcal{O}(\epsilon^2) \\
&\sim \epsilon\! \int dr \, \fr{d}{dr} \Big(r^{d-2} s(r) s'(r) \Big) - (d-3)\, \epsilon \!\int dr \, r^{d-3} s(r) s'(r) + \mathcal{O}(\epsilon^2). \nn
\ea
The first term is a total derivative and therefore does not contribute to the RT equation. The second term is down by a factor of $1/r$ compared to the contribution from the Einstein-Hilbert term \er{eq:inducedDet}. Since we are only interested in the leading large-$r$ profile, this term will not contribute either. We therefore conclude that the Gauss-Bonnet correction does not modify \er{eq:einmutd}, reproducing the result from \cite{Mezei:2016wfz}.

\section{Butterfly velocities for \texorpdfstring{$f$}{f}(Riemann) gravity}
\label{sec:butterfly}
In this section we derive general formulae for the butterfly velocities using the two holographic methods described in Section~\ref{sec:methods}, valid for all $f(\text{Riemann})$ theories. Before diving into the derivations, we will lay out a few useful definitions and notations and calculate a few basic quantities related to the background metric.

\subsection{Definitions, notations, and the spacetime}
As reviewed in Section~\ref{ssec:rev_shock}, the metric for the shockwave geometry is given by
\be \label{eq:shock}
ds^2 = 2A(uv) dudv +B(uv) dx^i dx^i -2A(uv) h(x) \d(u) du^2.
\ee
The background metric with $h = 0$ is simply the planar black hole \eqref{eq:bg_metric_kruskal}. We will denote the functions $A$, $B$ and their derivatives evaluated on the horizon at $u = 0$ by $A_n\equiv \partial^n A(uv)/\partial (uv)^n|_{u=0}$ and $B_n\equiv \partial^n B(uv)/\partial (uv)^n|_{u=0}$.  Throughout our calculations, we set $B_0=1$ by rescaling the $x^i$ coordinates.

The non-zero Christoffel symbols are
\begin{equation}
\begin{aligned}
    \T\G{v}{uu} =& -h\d'(u) +vA^{-1}A'h\d(u),\qu
    \T\G{u}{uu} = vA^{-1}A',\qu
    \T\G{v}{vv} = uA^{-1}A',\\
    \T\G{i}{uu} =&\, AB^{-1}h_j \d^{ij} \d(u),\qu
    \T\G{v}{ui} = -h_i \d(u),\qu
    \T\G{v}{ij} = -\fr{1}{2}vA^{-1}B'\d_{ij},\\
    \T\G{u}{ij} =& -\fr{1}{2}uA^{-1}B'\d_{ij},\qu
    \T\G{i}{ju} = \fr{1}{2}vB^{-1}B'\d^i_{j},\qu
    \T\G{i}{jv} = \fr{1}{2}uB^{-1}B'\d^i_{j},
\end{aligned}
\end{equation}
where $h_i\equiv \partial_i h(x)$, and we have dropped terms\footnote{\label{ft:delta}We can drop these terms at this stage because, as we will see, they will not be multiplied by any negative powers of $u$ in the equations of motion.  This allows us to use the identities $u \delta(u)=0$ and $u \delta'(u)=-\delta(u)$, viewed as equality of distributions.} containing $u \delta(u)$.  It will become clear in our later calculations that the second term $vA^{-1}A'h\d(u)$ in $\T\G{v}{uu}$ is unimportant because it always enters the equations of motion together with at least one power of $u$.

The non-zero components of the Riemann tensor are
\begin{subequations}
\label{eqs:riemanns}
\ba 
R_{uvuv} &= A' +uv A'' -uv A^{-1} A'^2,\\
R_{uivj} &= -\fr{1}{2}\d_{ij} \(B' +uv B'' -\fr{1}{2}uv B^{-1} B'^2\),\\
R_{uiuj} &= \fr{\d_{ij}}{2}v^2 \(A^{-1}A'B' +\fr{1}{2} B^{-1}B'^2 -B''\) +\(Ah_{ij} +\fr{\d_{ij}}{2} B' h\)\d(u), \\
R_{vivj} &= \fr{\d_{ij}}{2}u^2 \(A^{-1}A'B' +\fr{1}{2} B^{-1}B'^2 -B''\),\\
R_{ijkl} &= \fr{1}{2} uv A^{-1} B'^2 (\d_{il}\d_{jk} -\d_{ik}\d_{jl}),
\ea
\end{subequations}
where we have again discarded terms which vanish as a distribution. We see that the only correction due to $h$ is in $R_{uiuj}$, and the correction is linear in $h$.  The same statement applies for $\Tu R{ui}{vj}$. The fact that, given the specific ansatz \er{eq:shock}, these are the only possible non-zero components of the Riemann tensor is crucial in establishing our proof.

For notational convenience, we define the following tensors and use them to denote the corresponding values evaluated in the background solution and on the $u=0$ horizon:
\begin{equation}
\begin{aligned}
    C_{\r\s} \eq& \,\fr{\pa \mathcal{L}}{\pa g^{\r\s}}, \qu D^{\m\n\r\s} \eq \fr{\pa \mathcal{L}}{\pa R_{\m\n\r\s}}, \qu F^{\m\n\r\s}{}_{\l\t} \eq \fr{\pa^2 \mathcal{L}}{\pa R_{\m\n\r\s}\pa g^{\l\t}},\\
    G_{\m\n\r\s} \eq&\, \fr{\pa^2 \mathcal{L}}{\pa g^{\m\n}\pa g^{\r\s}}, \qu H^{\m\n\r\s\m'\n'\r'\s'} \eq \fr{\pa^2 \mathcal{L}}{\pa R_{\m\n\r\s}\pa R_{\m'\n'\r'\s'}}.
\end{aligned}
\end{equation}
For example, $C_{\r\s}$ denotes the value of $\pa \mathcal{L}/\pa g^{\r\s}$ in the background $h=0$ solution and on the $u=0$ horizon. Here, we have viewed the Lagrangian $\mathcal{L}$ as a function of $R_{\m\n\r\s}$ and $g^{\m\n}$, i.e., we lower all indices on the Riemann tensor and raise all indices on the metric and treat these two types of tensors as independent variables when taking derivatives. Throughout this paper, we define derivatives with respect to tensors such as $R_{\m\n\r\s}$ and $g^{\m\n}$ in the standard way; for example, $\pa \mathcal{L}/\pa R_{\m\n\r\s}$ has the Riemann symmetry and satisfies the identity $\d \mathcal{L} = \frac{\pa \mathcal{L}}{\pa R_{\m\n\r\s}} \d R_{\m\n\r\s}$.

Since the $i,j$ indices can only appear in the combination $\delta_{ij}$ for any background quantity, we can define the following notation where the transverse components are stripped away:
\bea
&& D^{aibj} \equiv D^{ab} \d^{ij},\qu \T F{aibj}{cd} \eq \T F{ab}{cd} \d^{ij},\qu \T F{abcd}{ij} \eq F^{(2)abcd}\delta_{ij}, \nn\\
&& H^{aibjckdl} \eq H^{(1)abcd} \d^{ij} \d^{kl} + \fr{H^{(2)abcd}}{2} (\d^{ik} \d^{jl} + \d^{il} \d^{jk}),\qu
H^{abcd}\eq H^{(1)abcd} + \fr{H^{(2)abcd}}{d-1},
\nn\\
&& H^{abcdeifj} \eq H^{(3)abcdef} \d^{ij}, \qu
H^{abcdijkl}  \eq H^{(4)abcd}(\d^{ik} \d^{jl}-\d^{il}\d^{jk}),\nn
\eea
where $a,b,c, \dots \in \{u,v\}$ and $i,j,k, \dots$ are the transverse directions.

\subsection{The shockwave method}

For general $f(\text{Riemann})$ theories, we start by writing down the equations of motion:
\be \label{eq:shockEOM}
E_{\mu \nu} \equiv \frac{2}{\sqrt{-g}}\frac{\delta S}{\delta g^{\mu \nu}}=E^{(1)}_{\m\n} + E^{(2)}_{\m\n} + E^{(3)}_{\m\n} + E^{(4)}_{\m\n},
\ee
with
\bea
\label{eq:EOM}
E^{(1)}_{\m\n} &=&- g_{\m\n} \mathcal{L}, \quad E^{(2)}_{\m\n}=2\fr{\pa \mathcal{L}}{\pa g^{\m\n}}, \nn\\
E^{(3)\m\n} &=& -2\fr{\pa \mathcal{L}}{\pa R_{\m\l\r\s}} \T R{\n}{\l\r\s} \bigg|_{\text{sym}(\m\n)}, \quad E^{(4)\m\n}=4\(\fr{\pa \mathcal{L}}{\pa R_{\m\r\n\s}}\)_{;\s;\r} \bigg|_{\text{sym}(\m\n)},
\eea
where the Lagrangian $\mathcal{L}$ is viewed as a function of $R_{\m\n\r\s}$ and $g^{\m\n}$.
Here the parenthesized numbers $(1),\dots,(4)$ merely label the various terms and do not have any physical meaning.

We view the shockwave spacetime \er{eq:shock} as a perturbation from the background geometry \er{eq:bg_metric_kruskal} with
\be
\d g_{uu} = -2A h\d(u).
\ee
The only non-zero component of the perturbation of the inverse metric is
\be
\d g^{vv} = 2A^{-1} h\d(u).
\ee
As we show in Appendix~\ref{app:lin}, the only component of the equations of motion $E^\n_\m$ that receives a non-vanishing perturbation from the shockwave is $E^v_u$, with
\be\la{eq:dEvu}
\delta E^v_u= g_{uv}\delta E^{vv}+E^{uv} \delta g_{uu}.
\ee
We now calculate these two terms separately.

For $\delta E^{vv}$, we have
\begin{subequations}\la{eq:evv}
\be
\d E^{vv} = \d E^{(1)vv} + \d E^{(2)vv} + \d E^{(3)vv} + \d E^{(4)vv},
\ee
where
\begin{align}
\begin{split}
    \d E^{(1)vv}=& - \d g^{vv} \mathcal{L} = -\fr{2\mathcal{L}}{A} h\d(u),
\end{split}
\\
\begin{split}
    \d E^{(2)vv} =&\, 4\d g^{vv} g^{uv} C_{uv} +2 (g^{uv})^2 G_{uuvv} \d g^{vv} + 8 (g^{uv})^2 \T F{uu}{uu} \d^{ij} \d R_{uiuj}\\
    =& \[\fr{8}{A^2} C_{uv} +\fr{4}{A^3} G_{uuvv} + \fr{8\T F{uu}{uu}}{A^2} \(A \partial_i\partial_i +\fr{d-1}{2} B'\)\] h\d(u),
\end{split}
\\
\begin{split}
    \d E^{(3)vv} =& - \left( 4 D^{vuvu}R_{vuvu}+4D^{viuj} R_{viuj} \right)\d g^{vv}-4 D^{viuj} g^{uv} \d R_{uiuj} \\
    &- 4 F^{viaj}{}_{vv} g^{uv} R_{uiaj} \d g^{vv}- 16 H^{uiujvkvl} g^{uv} R_{ukvl} \d R_{uiuj} \\
    =& \bigg[ -\fr{8A'}{A} D^{uvuv}+4(d-1) \fr{B'}{A} \(D^{uv}+F^{vv}{}_{vv}\) \\
    & - \(4 D^{uv}-8(d-1) H^{uuvv} B'\) \bigg(\partial_i\partial_i+\fr{d-1}{2} \fr{B'}{A} \bigg) \bigg] h\d(u),
\end{split}
\\
\begin{split}
    \d E^{(4)vv} =& \,4 \delta\left(\nabla_\r\nabla_\s\frac{\partial \mathcal{L}}{\partial R_{v\r v\s}} \right) \\
=&\, 4\T F{vivj}{vv} \T{\d g}{vv}{,j,i} +16 H^{ukulvivj} \d R_{ukul,j,i} \\
&+4 \d \T\G{v}{uj,i} D^{uivj} +4 \d \T\G{i}{uu,i} D^{vuvu} +4 \d \Gamma_{uu}^v D^{vuu\sigma}{}_{;\sigma}\\
=&\bigg[\fr{8}{A} \T F{vv}{vv} \partial_i\partial_i + 8 H^{(1)uuvv} \(2A\partial_i\partial_i\partial_j\partial_j +(d-1) B'\partial_i\partial_i\) \\
\quad& +8 H^{(2)uuvv} \(2A\partial_i\partial_i\partial_j\partial_j + B'\partial_i\partial_i \) - 4 D^{uv} \partial_i\partial_i  +4 D^{uvuv} \fr{A}{B} \partial_i\partial_i  \\
&-2(d-1) \frac{B'}{A} D^{uv} - \( 8\frac{A'}{A} +2 (d-1) \frac{B'}{B}  \)D^{uvuv} +8 \frac{A'}{A^2} F^{uvuv}{}_{uv} \\
&+  4(d-1)\frac{B'}{B^2} F^{(2)uvuv} -16\(2A''-\frac{A'^2}{A}\) H^{uvuvuvuv} \\
&-32(d-1) \( \frac{B'^2}{4B} - B'' \) H^{(3)uvuvuv} + 4(d-1)(d-2) \frac{B'^2}{A} H^{(4)uvuv} \bigg] h\d(u).
\end{split}
\end{align}
\end{subequations}
In arriving at this, it is important that we use the distributional identity $u\d'(u) =- \d(u)$ (see Footnote~\ref{ft:delta}). The $\d$-function sets $u=0$ so the quantities are all evaluated on the horizon. In deriving \er{eq:evv}, it is useful to note the following simplifying properties. First, in the background solution, every extra $v$-index downstairs (beyond those paired with a $u$-index downstairs or a $v$-index upstairs) costs a factor of $u$. Similarly, a single $i$-type index cannot contribute in the background solution since it must come paired with another such index to form a Kronecker delta.

To work out the second term on the right-hand side of \er{eq:dEvu}, we find an expression for $E^{uv}$ in the background solution on the horizon:
\begin{subequations}
\be
E^{uv}=E^{(1)uv}+E^{(2)uv}+E^{(3)uv}+E^{(4)uv},
\ee
where
\begin{align}
\begin{split}
    E^{(1)uv}=&-\fr{\mathcal{L}}{A},
\end{split}
\\
\begin{split}
    E^{(2)uv}=&\,\fr{2C_{uv}}{A^2},
\end{split}
\\
\begin{split}
    E^{(3)uv}=&\,2(d-1)\fr{B'}{A} D^{uv}-\fr{4A'}{A}D^{uvuv},
\end{split}
\\
\begin{split}
    E^{(4)uv}=&\, \fr{8A'}{A^2}F^{uvuv}{}_{uv}+4(d-1)B'F^{(2)uvuv}-16\(2A''-\fr{A'^2}{A} \)H^{uvuvuvuv} \\
    & +16(d-1)\bigg(2B''-\fr{B'^2}{2} \bigg)H^{(3)uvuvuv}+4(d-1)(d-2)\fr{B'^2}{A}H^{(4)uvuv} \\
    & +2\left(\fr{4A'}{A}+(d-1)B' \right)D^{uvuv}-2(d-1)\fr{B'}{A} D^{uv}.
\end{split}
\end{align}
\end{subequations}

Finally, collecting all the terms into \eqref{eq:dEvu} and plugging in the ansatz for $h(x)$ yields
\be
\lim_{r\to\infty} \lt.\frac{1}{4h(x)} \delta E^v_u\rt|_{h(x)\sim\frac{e^{-\mu r}}{r^{\#}}} \equiv \delta(u)f_\text{SW}(\mu)=0,
\ee
where we have taken the large-$r$ limit and neglected higher-order terms in $1/r$, and
\begin{multline}\label{eq:fSW}
f_{\text{SW}}(\mu) = \frac{{C}_{uv}}{A^2_0}+\bigg(\frac{2 A_1 }{A_0}+\fr{d-1}{2}B_1 \bigg) {D}^{uvuv}-(d-1){B_1 }{F}^{(2)uvuv} \\
-\fr{2A_1}{A_0^2}{F}^{uvuv}{}_{uv}+(d-1)\fr{B_1}{A_0^2}\({F}^{uu}{}_{uu}+{F}^{vv}{}_{vv}\) +\fr{{G}_{uuvv}}{A_0^3}\\
+(d-1)^2\frac{B_1^2}{A_0}{H}^{uuvv}+2(d-1)\bigg( B_1^2-4B_2\bigg){H}^{(3)uvuvuv} \\
-(d-1)(d-2)\frac{B_1^2 }{A_0}{H}^{(4)uvuv}+4\bigg(2A_2-\frac{A_1^2}{A_0} \bigg) {H}^{uvuvuvuv} \\
+\(-2{D}^{uv}+{A_0}{D}^{uvuv}+\frac{2 }{A_0}\left({F}^{uu}{}_{uu}+{F}^{vv}{}_{vv}\right) +4(d-1)B_1 {H}^{uuvv}\)\mu^2 \\
+ 4A_0 \Big({H}^{(1)uuvv}+{H}^{(2)uuvv}\Big) \mu^4
\end{multline}
is a function of quantities on the horizon through imposing $\delta(u)$. This is a quartic\footnote{It is quartic because we are considering $f($Riemann$)$ theories, which have only up to four derivatives acting on a single factor of the metric in the equations of motion \eqref{eq:EOM}.} equation for $\mu$ and a quadratic equation for $\mu^2$. The correct root is the one that is positive and continuously connected to the unperturbed Einstein gravity result given in Eq.~\eqref{eq:einmu}. We then extract the butterfly velocity $v_B$ from~\eqref{eq:vB}.

\subsection{The entanglement wedge method}
\label{sec:butterflyew}
Let us now move on to the entanglement wedge method. As reviewed in Section~\ref{ssec:EW}, our objective is to find the size of the smallest spherical boundary region whose entanglement wedge encloses a probe particle falling into the black hole.

We will work in the same coordinate system as used for the shockwave method. This will make it easier to see the matching with the shockwave result, but at the expense of making the time translation symmetry slightly less explicit. The metric is the planar black hole given by
\begin{equation} 
d s^{2} = 2A(u v)dudv+B(u v)dx^{i}dx^{i}.
\end{equation}
We would now like to derive the RT equation for a spherical boundary region in this background using~\eqref{eq:S_EE}. Before we proceed, let us make the following simplifying observations:
\begin{enumerate}
    \item Entanglement surfaces anchored to a single boundary can never penetrate the horizon, so we can choose to work in one of the exterior patches and exploit the time translation symmetry. Because of this symmetry, we only need to look for the RT surface rather than the HRT surface. This means we can restrict to the $u=-v$ hypersurface in order to get the spatial profile of the entanglement surface. This is the $t=0$ surface in the original $(t,z,x^i)$ coordinates.
    \item Since we are interested in the near-horizon limit, the butterfly velocity can be calculated by extremizing the entropy functional $S_{\text{EE}}$ with respect to a candidate RT surface defined by $uv=-\epsilon s(x)^2$ to linear order in $\epsilon$. Since $u=-v$, we can consider each factor of $u$ or $v$ as contributing a factor of $\sqrt{\epsilon}$.
    \item The entropy functional $S_{\text{EE}}$ given in Eq.~\er{eq:S_EE} is only accurate at second order in the extrinsic curvature $K$ and its derivatives, but this is sufficient to determine the butterfly velocity. In particular, higher-order terms in $K$ and its derivatives are suppressed by additional powers of $\e$, and can thus be neglected in the near-horizon limit.
    \item The $\varepsilon^{\lambda_1\lambda_2}$ term in Eq.~\er{eq:S_EE} vanishes when restricting to RT surfaces. To see this, go to a coordinate system where the time translation symmetry is manifest (so $\partial_t$ is the timelike Killing vector field), and note that $K_{\lambda\rho\sigma}$ vanishes if $\lambda = t$ but $\varepsilon^{\lambda_1\lambda_2}$ vanishes unless one of $\lambda_1$ and $\lambda_2$ is $t$. 
    \item We can write $K_{\lambda_{1} \rho_{1} \sigma_{1}} K_{\lambda_{2} \rho_{2} \sigma_{2}} n^{\lambda_{1} \lambda_{2}}=K_{2 \rho_{1} \sigma_{1}} K_{2 \rho_{2} \sigma_{2}}$, where `2' denotes the direction of the second normal vector (which is orthogonal to the $t$ direction), i.e., $K_{2 \rho \sigma}=h^\mu_\rho h^\nu_\sigma \nabla_\mu n^{(2)}_\nu$, where $h^\mu_\nu$ is the projector onto the codimension-2 surface. This is a simplification due to the first observation above: the extrinsic curvature $K_{1\rho\sigma}\equiv K_{t\rho\sigma}=0$ because of the time reflection symmetry at $t=0$. 
\end{enumerate}
Implementing these simplifications and writing $S_\text{EE}=2\pi\int d^{d-1}y\,\sqrt{\gamma}\mathcal{L}_\text{EE}$, we have
\begin{equation}
\begin{split}
    \mathcal{L}_\text{EE}=& -\frac{\partial \mathcal L}{\partial R_{\mu \rho \nu \sigma}} \varepsilon_{\mu \rho} \varepsilon_{\nu \sigma}\\
    &-\frac{\partial^{2} \mathcal L}{\partial R_{\mu_{1} \rho_{1} \nu_{1} \sigma_{1}} \partial R_{\mu_{2} \rho_{2} \nu_{2} \sigma_{2}}} 2 K_{2 \rho_{1} \sigma_{1}} K_{2 \rho_{2} \sigma_{2}} \left(n_{\mu_{1} \mu_{2}} n_{\nu_{1} \nu_{2}}+\varepsilon_{\mu_{1} \mu_{2}} \varepsilon_{\nu_{1} \nu_{2}}\right).
\end{split}
\end{equation}
We call the second term the extrinsic curvature term.

The non-zero components of the Riemann tensor in the background solution are again given by \er{eqs:riemanns} but with $h$ set to zero, i.e., without the shockwave. With our candidate entanglement surface defined on $uv=-\epsilon s(x)^2$,  the components of the two normals are then given by
\begin{equation}
    \begin{aligned}
    n^{(1)}_u &= \sqrt{\frac{v A(uv)}{2 u}}, \quad n^{(2)}_u =  \frac{-v}{\sqrt{2uv/A(uv)+4\epsilon^2s^2 s_js_j /B(uv)}}, \\
    n^{(1)}_v &= \sqrt{\frac{u A(uv)}{2 v}}, \quad n^{(2)}_v =  \frac{-u}{\sqrt{2uv/A(uv)+4\epsilon^2s^2 s_js_j /B(uv)}},\\
    n^{(1)}_i &= 0, \quad n^{(2)}_i=\frac{-2\epsilon ss_i}{\sqrt{2uv/A(uv)+4\epsilon^2s^2 s_js_j/B(uv)}},
\end{aligned}
\end{equation}
where $s_i$ stands for $\pa_i s(x)$. In deriving this, we used the fact that $t$ is a function of $u/v$, $n^{(1)}\sim dt$, and $n^{(2)}\sim df$ where $f=uv+\epsilon s^2$. 

Next, we need the following tensors, defined by
\begin{align}
n_{\mu\nu}&=-n_{\mu}^{(1)} n_{\nu}^{(1)}+n^{(2)}_\mu n^{(2)}_\nu, \\
\varepsilon_{\mu\nu}&=n_{\mu}^{(1)} n_{\nu}^{(2)}-n^{(2)}_\mu n^{(1)}_\nu.
\end{align}
To linear order in $\epsilon$, the non-zero components of $\varepsilon_{\mu\nu}$ are given by
\begin{equation}
    \begin{aligned}
    \varepsilon_{u v}&=A_0-\epsilon\left(s^2 A_1-{A_0^{2} s_{j} s_{j}}\right),\\
    \varepsilon_{u i}&=\sqrt{\frac{-\epsilon\, v}{u}} A_0 s_{i} +\mathcal{O}(\epsilon^{3 / 2}),\\
    \varepsilon_{v i}&=\sqrt{\frac{-\epsilon\, u}{v}} A_0 s_i+\mathcal{O}(\epsilon^{3 / 2}).
\end{aligned}
\end{equation}
It turns out that we will only need the $\mathcal{O}(1)$ term in $n_{\mu\nu}$, and the only non-zero component at this order is
\begin{equation}
    n_{uv}=A_0 + \mathcal{O}(\epsilon).
\end{equation}
We will also need the extrinsic curvatures. To leading order, the only non-zero component is
\begin{equation}
K_{2 i j}=\sqrt{\frac{-\epsilon}{2 A_0}}\left(-B_1\delta_{ij}s+A_0s_{ij}\right).
\end{equation}

We can now derive the contributions to $S_\text{EE}$ at linear order in $\epsilon$. For any quantity $X$, we denote the linear order coefficient in a Taylor expansion in $\epsilon$ by $\Delta X$. Then $\Delta S_\text{EE}=2\pi\int d^{d-1}y\,\Delta(\sqrt{\gamma}\mathcal{L}_\text{EE})$ is given by
\begin{align}
    \Delta(\sqrt{\gamma}\mathcal{L}_\text{EE})&\equiv(\Delta\sqrt{\gamma})\bar{\mathcal{L}}_{\text{EE}} +\sqrt{\bar\gamma}\(\Delta \mathcal{L}_\text{EE}^{(1)}+\Delta \mathcal{L}_\text{EE}^{(2)}+\Delta \mathcal{L}_\text{EE}^{(3)}\),\\
    \bar{\mathcal{L}}_{\text{EE}}&=- {D}^{\mu\rho\nu\sigma}\bar\varepsilon_{\mu\rho}\bar\varepsilon_{\nu\sigma},\\
    \Delta\mathcal{L}_\text{EE}^{(1)}&=-\bar\varepsilon_{\mu\rho}\bar\varepsilon_{\nu\sigma}\Delta\( \frac{\partial\mathcal{L}}{\partial R_{\mu\rho\nu\sigma}}\),\\
    \Delta\mathcal{L}_\text{EE}^{(2)}&=-\Delta(\varepsilon_{\mu\rho}\varepsilon_{\nu\sigma}) {D}^{\mu\rho\nu\sigma},\\
    \Delta\mathcal{L}_\text{EE}^{(3)}&=-\Delta\left(2K_{2\rho_1\sigma_1}K_{2\rho_2\sigma_2}\right)\left(\bar n_{\mu_{1} \mu_{2}} \bar n_{\nu_{1} \nu_{2}}+\bar \varepsilon_{\mu_{1} \mu_{2}} \bar\varepsilon_{\nu_{1} \nu_{2}}\right){H}^{\mu_1\rho_1\nu_1\sigma_1\mu_2\rho_2\nu_2\sigma_2},
\end{align}
where the barred quantities are evaluated on the horizon at $\epsilon=0$. Note that quantities such as ${D}^{\mu\rho\nu\sigma}$ and ${H}^{\mu_1\rho_1\nu_1\sigma_1\mu_2\rho_2\nu_2\sigma_2}$ do not need to have bars because they are already defined to be evaluated on the horizon. The last piece is the only contribution from the extrinsic curvature term of $S_{\text{EE}}$ since $K_{2ij}= \mathcal{O}(\sqrt \epsilon)$.

The determinant of the induced metric is given by
\begin{equation}
    \sqrt{\gamma}=1-{\epsilon}\left({A_0}s_js_j+\frac{d-1}{2}B_1 s^2\right)+\mathcal{O}(\epsilon^2),
\end{equation}
which can be derived by substituting $uv=-\epsilon s(x)^2$ into the metric and expanding the identity $\det \exp M = \exp \operatorname{Tr} M$ to linear order. Then
\begin{equation}
    \fr{1}{4A_0^2}(\Delta\sqrt{\gamma})\bar{\mathcal{L}}_{\text{EE}}=\left({A_0s_j s_j}+\frac{d-1}{2}B_1 s^2\right)D^{uvuv},
\end{equation}
where we have used the fact that only $\varepsilon_{uv}\ne0$ at zeroth order.

For the next term, we need
\begin{align}
    \Delta\( \frac{\partial\mathcal{L}}{\partial R_{\mu\rho\nu\sigma}}\)
    &={H}^{\mu\rho\nu\sigma\mu^\prime\rho^\prime\nu^\prime\sigma^\prime}\Delta R_{\mu^\prime\rho^\prime\nu^\prime\sigma^\prime}
    +2{F}^{\mu\rho\nu\sigma}{}_{uv}\Delta g^{uv}
    +{F}^{\mu\rho\nu\sigma}{}_{ij}\Delta g^{ij},\\
    g^{uv}&=\frac{1}{A(-\epsilon s^2)}=\frac{1}{A_0}+\epsilon\frac{A_1}{A_0^2}s^2+\mathcal{O}(\epsilon^2),\\
    g^{ij}&=\frac{\delta^{ij}}{B(-\epsilon s^2)}=\delta^{ij}\left(1+\epsilon {B_1 s^2}+\mathcal{O}(\epsilon^2)\right).
\end{align}
We also notice that $H$ vanishes if the numbers of lower $u$ and $v$ indices do not match (each upper $u$ is considered one lower $v$ and vice versa). Then we have 
\begin{multline}
    \fr{1}{4A_0^2}\sqrt{\bar\gamma}\Delta\mathcal{L}_\text{EE}^{(1)}=\bigg[4\(2A_2-\frac{A_1^2}{A_0}\) H^{uvuvuvuv}
    -4 \(2B_2-\frac{1}{2}B_1^2\) \delta_{ij}  H^{uvuvuivj}\\
     +\frac{B_1^2}{2A_0}\(\delta_{il}\delta_{jk}-\delta_{ik}\delta_{jl}\) H^{uvuvijkl}
    -\frac{2A_1}{A_0^2} F^{uvuv}{}_{uv}
    -{B_1}\delta^{ij}{F}^{uvuv}{}_{ij} \bigg ] s^2.
\end{multline}
Using the expressions for $\varepsilon_{\m\n}$ above, the third term is simply given by
\begin{equation}
    \fr{1}{4A_0^2}\sqrt{\bar\gamma}\Delta\mathcal{L}_\text{EE}^{(2)}=2D^{uvuv}\left(\fr{A_1}{A_0}s^2-{A_0}s_js_j\right)-2 D^{uivj}{s_is_j}.
\end{equation}
In the last term $\Delta \mathcal{L}_\text{EE}^{(3)}$, we notice that only $K_{2ij}$ has low enough order in $\epsilon$ to contribute, so we have
\begin{equation}
    \fr{1}{4A_0^2}\sqrt{\bar\gamma}\Delta\mathcal{L}_\text{EE}^{(3)}=  H^{uiujvkvl}
    \left[\fr{B_1^2}{A_0}\delta_{ij}\delta_{kl} s^2 - 2B_1 {s(s_{ij}\delta_{kl}+s_{kl}\delta_{ij})}+4A_0{s_{ij}s_{kl}}\right].
\end{equation}
Finally, putting everything together, the total contribution to the entropy functional at linear order in $\e$ is given by
\be
\begin{aligned}
& \fr{1}{4A_0^2} \Delta(\sqrt{\gamma}\mathcal{L}_\text{EE}) \\= &\left[\left( \frac{2A_1}{A_0}  + \fr{d-1}{2}{B_1}\right){D}^{uvuv} - {B_1}\delta^{ij} {F}^{uvuv}{}_{ij} - \frac{2A_1}{A_0^2} {F}^{uvuv}{}_{uv} \right.\\
& \qqu + \frac{B_1^2}{A_0}\delta_{ij}\delta_{kl} {H}^{uiujvkvl} + 2\left({B_1^2} - 4B_2 \right) \delta_{ij} {H}^{uvuvuivj} \\
& \qqu \left. + \frac{B_1^2}{2A_0}\left(\delta_{il}\delta_{jk}-\delta_{ik}\delta_{jl}\right) {H}^{uvuvijkl}
  + 4\left(2A_2- \frac{A_1^2}{A_0} \right) {H}^{uvuvuvuv} \right]s^2 \\
& - \left({A_0}  D^{uvuv}\delta_{ij}   +2 D^{uivj} \right) s_is_j
 -2{H}^{uiujvkvl}
    \Big[B_1 {s\(s_{ij}\delta_{kl}+s_{kl}\delta_{ij}\)}-2A_0{s_{ij}s_{kl}}\Big].
\end{aligned}
\ee

To obtain the butterfly velocity, we vary $\Delta S_\text{EE}=2\pi\int d^{d-1}y\,\Delta(\sqrt{\gamma}\mathcal{L}_\text{EE})$ with respect to $s(x)$ and then substitute our ansatz $s(x) \sim \frac{e^{\wtd\m r}}{r^\#}$ from \er{eq:S_ansatz}, keeping only leading terms in $1/r$. It is not hard to see that the number of $x^i$-derivatives on $s$ will be the number of factors of $\wtd\mu$. From this, we obtain a polynomial equation for $\wtd\mu$:
\be
\lim_{r\to \infty}\left.\frac{1}{16\pi A_0^2 s(x)}\frac{\delta\left( \Delta S_\text{EE}\right)}{\delta s(x)}\right|_{s(x)\sim \frac{e^{\wtd\mu r}}{r^\#}}\equiv f_{\text{EE}}(\wtd\mu)=0,
\ee
where
\begin{multline}\label{eq:fEE}
    f_{\text{EE}}(\wtd\mu)= \bigg(\fr{2A_1}{A_0} + \fr{d-1}{2} B_1 \bigg) D^{uvuv} - (d-1)B_1 F^{(2)uvuv} - \fr{2A_1}{A_0^{2}} F^{uvuv}{}_{uv} \\
    + (d-1)^2\frac{B_1^2}{A_0} H^{uuvv} 
      + 2(d-1) \bigg({B_1^2} - 4 B_2 \bigg)  H^{(3)uvuvuv} \\- (d-1)(d-2)\frac{B_1^2}{A_0} H^{(4)uvuv} + 4 \left(2A_2 - \frac{A_1^2}{A_0} \right) H^{uvuvuvuv} 
      \\+ \bigg( 2  D^{uv} + A_0  D^{uvuv} - 4 B_1 (d-1)  H^{uuvv}\bigg) \wtd\m^2 
     + 4 A_0 \Big(H^{(1)uuvv} +  H^{(2)uuvv} \Big) \wtd\m^4.
\end{multline}
Notice that all coefficients only involve quantities evaluated on the horizon; this is true as well in the shockwave calculation, where it is enforced by the presence of $\d(u)$. Similar to the shockwave result \eqref{eq:fSW}, this is a quartic equation for $\wtd \m$ and a quadratic equation for $\wtd \m^2$, from which we choose the positive root continuously connected to the result for Einstein gravity \eqref{eq:einmu}. The butterfly velocity $\wtd v_B$ is then obtained from $\wtd\m$ using \er{eq:vBtd}. \\

Before proceeding to show that the two butterfly velocities we have derived agree, let us pause for a moment and use the results from this and the previous subsection in an explicit example. Consider the following four-derivative correction to Einstein gravity:
\be
\mathcal{L} \supset R_{\mu\nu\rho\sigma}R^{\mu\nu\rho\sigma} = R_{\mu\nu\rho\sigma}R_{\mu^\prime\nu^\prime\rho^\prime\sigma^\prime}g^{\mu\mu^\prime}g^{\nu\nu^\prime}g^{\rho\rho^{\prime}}g^{\sigma\sigma^\prime}.
\ee
The non-vanishing tensor components are given by
\bea
    && C_{uv}=\left.2(R_{uvuv})^2(g^{uv})^3+4R_{uivj}R_{vkul}g^{uv}\d^{ik}\d^{jl}\right|_{u=0}=\frac{8A_1^2}{A_0^3}+2(d-1)\frac{B_1^2}{A_0},\nn\\
    && G_{uuvv}=\left.8R_{uivj}R_{ukvl}\d^{ik}\d^{jl}+8R_{uvuv}R_{uvvu}(g^{uv})^2\right|_{u=0}=2(d-1)B_1^2-\frac{8A_1^2}{A_0^2},\nn\\
    && D^{uvuv}=\left.2R^{uvuv}\right|_{u=0}=\frac{2A_1}{A_0^4} ,\qu
    D^{uivj}=2 R^{uivj}|_{u=0}=-\frac{B_1}{A_0^2} \delta^{ij},\nn\\
    &&  F^{uvuv}{}_{uv}=\lt.\frac{4}{A_0^{3}}R_{uvuv}\rt|_{u=0}=\frac{4A_1}{A_0^3},\qu F^{uivj}{}_{uv}=\left.2R^{uivj}g_{uv}\right|_{u=0}=-\frac{B_1}{A_0}\d^{ij},\nn \\
    && F^{uiuj}{}_{uu}=F^{vivj}{}_{vv}=4R^{uivj}g_{uv}|_{u=0}=-\frac{2B_1}{A_0}\d^{ij},\nn\\
    &&  H^{uvuvuvuv}=\frac{1}{2}(g^{uv})^4|_{u=0}=\frac{1}{2A_0^4},\qu     H^{uiujvkvl}=\frac{1}{4A_0^2}(\delta^{ik}\delta^{jl}+\delta^{il}\delta^{jk}).
\eea
Substituting these expressions into either \er{eq:fSW} or \er{eq:fEE} reproduces the result in \cite{Mezei:2016wfz}.

\section{Equivalence of the two butterfly velocities} \label{sec:proof}
In Section~\ref{sec:butterfly}, we derived general expressions for the butterfly velocities from two distinct holographic calculations --- the shockwave method and the entanglement wedge method. More specifically, we have obtained polynomial equations for the parameters $\m$ and $\wtd \m$ given by $f_{\text{SW}}(\mu)=0$ and $f_{\text{EE}}(\wtd\mu)=0$, respectively. 

In both cases, to solve for the value of $\mu$ (or $\wtd\mu$) we treat the higher-derivative couplings perturbatively and choose the positive root that is continuously connected to the value in Einstein gravity. Recalling that the butterfly velocities are related to these parameters via \er{eq:vB} and \eqref{eq:vBtd}, it then suffices to prove that $f_{\text{SW}}$ and $f_{\text{EE}}$ are the same function.

With $f_\text{SW}$ given in \eqref{eq:fSW} and $f_\text{EE}$ given in \eqref{eq:fEE}, the two functions are the same if the following two equations hold in the background solution and on the $u=0$ horizon:
\begin{subequations}
\setlength{\abovedisplayskip}{0pt}
\ba
    & C_{uv} + g^{uv} G_{uuvv} - 2 \d^{ij} R_{uivj} \left( F^{uu}{}_{uu} + F^{vv}{}_{vv} \right) = 0 \label{eq:rel1}, \\
    & (F^{uu}{}_{uu} + F^{vv}{}_{vv}) - 2 g_{uv} D^{uv} - 8 g_{uv} \d^{ij} R_{uivj} H^{uuvv} = 0 \label{eq:rel2}.
\ea
\end{subequations}
In the rest of this section, we use `on the background' to mean `in the background solution and on the $u=0$ horizon'.  In writing the above two equations, we have used the fact that $g^{uv} = A_0^{-1}$ and $\d^{ij} R_{uivj} = -\fr{1}{2}(d-1)B_1$ on the background. We will refer to \er{eq:rel1} and \er{eq:rel2} as the first and second relations, respectively. 

We will now prove the above relations for any given choice of Lagrangian involving contractions of Riemann tensors. To that end, it is useful to think of the terms in \er{eq:rel1} and \er{eq:rel2} as differential operators acting on the Lagrangian $\mathcal{L}$. For example, $C_{uv}$ can be thought of as the operator
\be
\widehat{C}_{uv} = \fr{\pa}{\pa g^{uv}}
\ee
acting on the Lagrangian $\mathcal{L}$ with the result evaluated on the background. A useful quantity to define is
\be
\wtd R_{ab} = \frac{1}{d-1} \d^{ij} R_{aibj},
\ee
which we use to rewrite
\be
\d_{kl} \fr{\pa}{\pa R_{akbl}} = \fr{1}{4} \fr{\pa}{\pa \wtd R_{ab}},
\ee
where the factor $1/4$ is a symmetry factor from the Riemann tensor. For example, when we act $\pa/\pa \wtd R_{uu}$ on a function of $R_{\m\r\n\s}$ such as $\mathcal L$, we take the derivative with respect to $\wtd R_{uu}$ while holding the traceless part of $R_{uiuj}$ fixed.  Now we can rewrite \er{eq:rel1} and \er{eq:rel2} by defining two operators
\begin{subequations}\label{eq:ops}
\ba 
\widehat O_1 &= \fr{\pa}{\pa g^{uv}} + g^{uv} \fr{\pa^2}{\pa g^{uu}\pa g^{vv}} - \fr{1}{2} \wtd R_{uv} \( \fr{\pa^2}{\pa \wtd R_{uu}\pa g^{uu}} +(u\to v)\), \\
\widehat O_2 &=
\wtd R_{uv} \fr{\pa}{\pa \wtd R_{uv}} + \wtd R_{uv}^2 \fr{\pa^2}{\pa \wtd R_{uu} \pa \wtd R_{vv}} -\fr{1}{2} g^{uv} \wtd R_{uv} \left(\fr{\pa^2}{\pa \wtd R_{uu} \pa g^{uu}} + (u \to v) \right).
\ea
\end{subequations}
Our goal is then to prove that
\be \label{eq:gen_op_eq}
\widehat{O}_i \mathcal{L} = 0, \qqu i = 1,2
\ee
on the background.  This will be the goal of the remainder of this section. 

For any Lagrangian composed of a covariant combination of an arbitrary number of the Riemann tensor and inverse metric, we expand it by decomposing the sum over any dummy index into two sums, one over $\{u,v\}$ and the other over the $x^i$ directions.  This can be written in the following schematic form
\be\la{eq:lsum}
\mathcal{L} = \sum L,
\ee
where $L$ is an object of the form
\be\label{eq:lgen}
L = g^{AA} \cd g^{AA} R_{AAAA} \cd R_{AAAA} R_{AIAI} \cd R_{AIAI} \mathcal{X}^{I\cd I}_{A\cd A}.
\ee
Here, $A$ denotes any $a$-type index labelling either $u$ or $v$, $I$ denotes any $i$-type index, and the tensor $\mathcal{X}$ is a product of any number of inverse metric and Riemann tensor components not explicitly shown in \er{eq:lgen}, i.e., $g^{II}$, $g^{AI}$, $R_{IIII}$, $R_{AIII}$, $R_{AAII}$, and $R_{AAAI}$.  Different $A$ (or $I$) indices may specialize to different $a$-type (or $i$-type) indices.  As a concrete example, \er{eq:lsum} for $\mathcal L=g^{\m\n} g^{\r\s} R_{\m\r\n\s}$ can be written as $\mathcal L = g^{ab} g^{cd} R_{acbd}+g^{ab} g^{ij} R_{aibj}+\cd$ where the first term $g^{ab} g^{cd} R_{acbd}$ is of the form $g^{AA} g^{AA} R_{AAAA}$ and the second term $g^{ab} g^{ij} R_{aibj}$ is of the form $g^{AA} R_{AIAI} g^{II}$, with $a,b,c, \dots \in \{u,v\}$ and $i,j,k,\dots$ labelling transverse coordinates. Notice that all $a$-type and $i$-type indices are contracted.

As we are only interested in $\widehat O_i \mathcal L$ on the background, we may simplify \er{eq:lgen} significantly by dropping those terms that vanish eventually.  In particular, $g^{AI}$, $R_{IIII}$, $R_{AIII}$, $R_{AAII}$, and $R_{AAAI}$ all vanish on the background.\footnote{This can be verified by setting $h=0$ and $u=0$ in \er{eqs:riemanns}.}  As the derivatives in $\widehat O_i$ do not involve these components, if $L$ in \er{eq:lgen} contains any of these components, it would vanish after acting with $\widehat O_i$ and evaluating on the background.  Therefore, we can restrict $L$ to those that do not contain any of these components.  Similarly, the traceless part $R_{aibj}-\wtd R_{ab} \d_{ij}$ of $R_{aibj}$ vanishes on the background, and as the derivatives in $\widehat O_i$ do not involve this traceless part, we can restrict $L$ to those that do not contain the traceless part, and thus we may replace all instances of $R_{AIAI}$ in \er{eq:lgen} with $\wtd{R}_{AA}$.
Therefore, we replace \er{eq:lgen} with
\be
L = g^{AA} \cd g^{AA} R_{AAAA} \cd R_{AAAA} \wtd R_{AA} \cd \wtd R_{AA},
\ee
up to a multiplicative constant that we do not need to keep track of.

Now define \textit{index loops} by connecting the two indices of $g^{ab}$, the two indices of $\wtd R_{ab}$, the first two indices of $R_{abcd}$, and its last two indices.  For example, the term $g^{ab} \wtd R_{ab}$ has a single (index) loop. In general, $L$ contains one or more loops, and the two antisymmetric pairs of indices in any $R_{abcd} = R_{[ab][cd]}$ need not be part of the same loop.  In order for a loop not to vanish on the background, it must consist of alternating $u$ and $v$: either $(uvuv\cd uv)$ or $(vuvu \cd vu)$. For example, $g^{ab}g^{cd}g^{ef} R_{afbc}\wtd{R}_{de}$ consists of a single loop $(abcdef)$, with non-vanishing contributions on the background
\be
g^{uv}g^{uv}g^{uv} R_{uvvu}\wtd{R}_{vu} + (u\leftrightarrow v),
\ee
while $g^{ab}g^{cd}g^{ef} R_{adbc}\wtd{R}_{ef}$ has the two loops $(abcd)$ and $(ef)$, with non-vanishing contributions on the background
\be
g^{uv}g^{uv}g^{uv} R_{uvvu}\wtd{R}_{uv} + g^{uv}g^{uv}g^{vu} R_{uvvu}\wtd{R}_{vu} + (u\leftrightarrow v).
\ee

It turns out that it is sufficient to prove $\widehat{O}_i L =0$ for $L$ made of a single loop, because even for $L$ made of multiple loops, $\widehat{O}_{i}$ must act entirely on a single loop to have a chance to be non-trivial: in particular, if we act the two derivatives in any second-derivative term of \er{eq:ops} --- such as $\pa^2/\pa \wtd R_{uu} \pa g^{uu}$ --- on two different loops, one of the two loops would have to contain an extra factor of $g^{uu}$, $\wtd R_{vv}$, $R_{vvab}$, or $R_{abvv}$, thus vanishing on the background.\footnote{For $\wtd R_{vv}$ this is because it is proportional to $u^2$, and thus vanishes on the horizon.}

Therefore, from now on we consider a general $L$ made of a single loop.  It may be written as
\be\label{eq:sing_loop}
L = g^{a_1b_1} g^{a_2b_2} \cd g^{a_nb_n} \mathcal{T}_{a_1b_1a_2b_2\cd a_nb_n}
\ee
where $n$ is the number of $g^{ab}$ factors and the tensor $\mathcal{T}$ is a product of a suitable number of $R_{abcd}$ and $\wtd R_{ab}$.\footnote{Although the two antisymmetric pairs of indices in $R_{abcd}$ need not be part of the same loop, this does not affect our analysis because $\widehat{O}_i$ does not involve $R_{abcd}$ at all; there is no harm in including the other antisymmetric pair of indices in $L$ even if they are not in the same loop.}  Our goal is thus to prove
\be
\widehat{O}_i L = 0, \qqu  i = 1,2,
\ee
on the background for any $L$ of the form \er{eq:sing_loop}. The general statement \er{eq:gen_op_eq} then follows.

Before proceeding, let us introduce some useful terminology. For simplicity, we rename the $g^{ab}$ factors appearing in \er{eq:sing_loop} so that the loop is precisely $(a_1 b_1 a_2 b_2 \cd a_n b_n)$. For any neighboring pair of inverse metrics $g^{a_kb_k}$, $g^{a_{k+1}b_{k+1}}$ (where $k=1,2,\cd,n$ and $a_{n+1}\eq a_1$, $b_{n+1}\eq b_1$), the $b_k, a_{k+1}$ indices are either (1) contracted with some $R_{b_k a_{k+1}cd}$ or $R_{cdb_k a_{k+1}}$, or (2) contracted with $\wtd R_{b_k a_{k+1}}$. In the first case, we say that there is an $R$-contraction between $g^{a_kb_k}$ and $g^{a_{k+1}b_{k+1}}$, while in the second case, we say that there is an $\wtd R$-contraction between $g^{a_kb_k}$ and $g^{a_{k+1}b_{k+1}}$. More generally, for any $k\leq l$ we say that there is an $R$-contraction between $g^{a_kb_k}$ and $g^{a_lb_l}$ if there is an $R$-contraction between any neighboring pair among $g^{a_kb_k}, g^{a_{k+1}b_{k+1}}, \cd, g^{a_{l}b_{l}}$, and we say that there is an $R$-contraction \textit{not} between $g^{a_kb_k}$ and $g^{a_lb_l}$ if there is an $R$-contraction between any of the other neighboring pairs (i.e., among $g^{a_1b_1}, \cd, g^{a_kb_k}$ or $g^{a_{l+1}b_{l+1}}, \cd, g^{a_{n}b_{n}}$).  Similar statements apply for $\wtd R$-contractions.  Note that the number of $R$-contractions and $\wtd R$-contractions add up to $n$. As an example, in the loop
$g^{a_1b_1}g^{a_2b_2}g^{a_3b_3}R_{a_1b_3b_1a_2}\wtd R_{b_2a_3}$, there is an $R$-contraction between $g^{a_3b_3}$ and $g^{a_1b_1}$, as well as between $g^{a_1b_1}$ and $g^{a_2b_2}$; and there is an $\wtd R$-contraction between $g^{a_2b_2}$ and $g^{a_3b_3}$.

\paragraph{First relation} We now prove the first relation
\be
\widehat O_1 L=0
\ee
on the background for any $L$ of the form \er{eq:sing_loop}.

First, consider $\widehat O_1^{(1)} L$ where
\be
\widehat O_1^{(1)} \equiv \fr{\pa}{\pa g^{uv}}.
\ee
On the background, we have
\be\label{eq:o1l}
\widehat O_1^{(1)} L = \widehat O_1^{(1)} L_1^{(1)}
\ee
where $L_1^{(1)}$ is the sum of the two terms in \er{eq:sing_loop} where the loop $(a_1b_1a_2b_2\cd a_n b_n)$ consists of alternating $u$ and $v$: either $(uvuv\cd uv)$ or $(vuvu \cd vu)$.  These two terms differ by a factor of $(-1)^m$ where $m$ is the total number of $R$-contractions,\footnote{Note that $m$ need not be an even integer because the two antisymmetric pairs of indices in a Riemann tensor can be in different loops.} because $\wtd R_{uv}=\wtd R_{vu}$ but exchanging $u$ and $v$ in each $R$-contraction (i.e., each antisymmetric pair of indices in $R_{abcd}$) costs a minus sign.  In other words,
\begin{equation}
\label{eq:L11}
\begin{aligned}
    L_1^{(1)} &= g^{uv}\cd g^{uv} \mathcal{T}_{uv\cd uv} + g^{vu}\cd g^{vu} \mathcal{T}_{vu\cd vu}\\
    &= \[1+ (-1)^m\] (g^{uv})^n \mathcal{T}_{uv\cd uv}
\end{aligned}
\end{equation}
and
\be
\widehat O_1^{(1)} L = \fr{1}{2} n g_{uv} L_1^{(1)},
\ee
where the factor of $1/2$ comes from the symmetry of $g^{ab}$.

Second, consider $\widehat O_1^{(2)} L$ where
\be
\widehat O_1^{(2)} \equiv g^{uv} \fr{\pa^2}{\pa g^{uu}\pa g^{vv}}.
\ee
On the background, we have
\be\label{eq:o2l}
\widehat O_1^{(2)} L = \widehat O_1^{(2)} L_1^{(2)}
\ee
where $L_1^{(2)}$ is the sum of all terms in \er{eq:sing_loop} where the loop $(a_1b_1a_2b_2\cd a_nb_n)$ is alternating except for two `defects' at $g^{a_kb_k}=g^{uu}$ and $g^{a_lb_l}=g^{vv}$, for any $k\neq l$.  The two derivatives in $\widehat O_1^{(2)}$ act precisely on these two defects.  If $k<l$, compared to the alternating loop $(uvuv\cd uv)$ we are exchanging $u$ and $v$ in all $R$- and $\wtd R$-contractions between $g^{a_kb_k}$ and $g^{a_lb_l}$. Since each $R$-contraction costs a minus sign and each $\wtd R$-contraction gives a plus sign, such a loop contributes
\be
(-1)^{s_{kl}} g^{uu}g^{vv}(g^{uv})^{n-2} \mathcal{T}_{uv\cd uv}
\ee
to $L_1^{(2)}$, where $s_{kl}$ (sometimes also written as $s_{k,l}$) is defined to be the number of $R$-contractions between $g^{a_kb_k}$ and $g^{a_lb_l}$.  By definition we have $s_{kl}=s_{lk}$ and $s_{kk}=0$, with no summation implied.

If $k>l$, compared to the alternating loop $(uvuv\cd uv)$ we are exchanging $u$ and $v$ in all $R$- and $\wtd R$-contractions \textit{not} between $g^{a_kb_k}$ and $g^{a_lb_l}$.  Since there is a total of $m$ $R$-contractions and thus the number of $R$-contractions \textit{not} between $g^{a_kb_k}$ and $g^{a_lb_l}$ is $m-s_{kl}$, such a loop contributes
\be
(-1)^{m-s_{kl}} g^{uu}g^{vv}(g^{uv})^{n-2} \mathcal{T}_{uv\cd uv}
\ee
to $L_1^{(2)}$.

Combining the above two cases, we find
\begin{equation}
\begin{aligned}
    L_1^{(2)} &= \sum_{k<l} (-1)^{s_{kl}} g^{uu}g^{vv}(g^{uv})^{n-2} \mathcal{T}_{uv\cd uv} + \sum_{k>l} (-1)^{m-s_{kl}} g^{uu}g^{vv}(g^{uv})^{n-2} \mathcal{T}_{uv\cd uv}\\
    &= \[1+ (-1)^m\] \sum_{k<l} (-1)^{s_{kl}} g^{uu}g^{vv}(g^{uv})^{n-2} \mathcal{T}_{uv\cd uv}
\end{aligned}
\end{equation}
and
\be
\widehat O_1^{(2)} L = \[1+ (-1)^m\] \sum_{k<l} (-1)^{s_{kl}} (g^{uv})^{n-1} \mathcal{T}_{uv\cd uv} = \sum_{k<l} (-1)^{s_{kl}} g_{uv} L_1^{(1)}.
\ee

Third, consider $\widehat O_1^{(3)} L$ where
\be
\widehat O_1^{(3)} \equiv -\fr{1}{2} \wtd R_{uv} \fr{\pa^2}{\pa \wtd R_{uu}\pa g^{uu}}.
\ee
On the background, we have
\be
\widehat O_1^{(3)} L = \widehat O_1^{(3)} L_1^{(3)}
\ee
where $L_1^{(3)}$ is the sum of all terms in \er{eq:sing_loop} where the loop $(a_1b_1a_2b_2\cd a_nb_n)$ is alternating except for two `defects' at $g^{a_kb_k}=g^{uu}$ and $\wtd R_{b_la_{l+1}}=\wtd R_{uu}$, for any $k$, $l$, whether or not they are equal.  If $k\leq l$, compared to the alternating loop $(uvuv\cd uv)$ we are exchanging $u$ and $v$ in all $R$- and $\wtd R$-contractions between $g^{a_kb_k}$ and $g^{a_lb_l}$. Such a loop contributes
\be\label{eq:l3c}
\fr{(-1)^{s_{kl}}+(-1)^{s_{k,l+1}}}{2} g^{uu}(g^{uv})^{n-1} \fr{\wtd R_{uu}}{\wtd R_{uv}} \mathcal{T}_{uv\cd uv}
\ee
to $L_1^{(3)}$. This expression is nice because it applies to any $l$ satisfying $k\leq l$, whether or not there is actually an $\wtd R$-contraction between $g^{a_lb_l}$ and $g^{a_{l+1}b_{l+1}}$. If there is, we have $s_{k,l+1}=s_{kl}$ and \er{eq:l3c} gives the correct contribution.  If not, there must be an $R$-contraction instead between $g^{a_lb_l}$ and $g^{a_{l+1}b_{l+1}}$, so we find $s_{k,l+1} = s_{kl} + 1$ and \er{eq:l3c} vanishes.

If $k>l$, compared to the alternating loop $(uvuv\cd uv)$ we are exchanging $u$ and $v$ in all $R$- and $\wtd R$-contractions \textit{not} between $g^{a_kb_k}$ and $g^{a_{l+1}b_{l+1}}$.  Such a loop contributes
\be
\fr{(-1)^{m-s_{kl}}+(-1)^{m-s_{k,l+1}}}{2} g^{uu}(g^{uv})^{n-1} \fr{\wtd R_{uu}}{\wtd R_{uv}} \mathcal{T}_{uv\cd uv}
\ee
to $L_1^{(3)}$.  Again, this expression vanishes if there is actually an $R$-contraction between $g^{a_lb_l}$ and $g^{a_{l+1}b_{l+1}}$.

Combining the above two cases, we find
\ba
    L_1^{(3)} &= \[ \sum_{k\leq l} \fr{(-1)^{s_{kl}}+(-1)^{s_{k,l+1}}}{2} + \sum_{k>l} \fr{(-1)^{m-s_{kl}}+(-1)^{m-s_{k,l+1}}}{2} \] g^{uu}(g^{uv})^{n-1} \fr{\wtd R_{uu}}{\wtd R_{uv}} \mathcal{T}_{uv\cd uv} \nn \\
    &= \[1+ (-1)^m\] \[ \fr{n}{2} + \sum_{k<l} (-1)^{s_{kl}} \] g^{uu}(g^{uv})^{n-1} \fr{\wtd R_{uu}}{\wtd R_{uv}} \mathcal{T}_{uv\cd uv}
\ea
and
\begin{equation}\label{eq:o3l}
\begin{aligned}
    \widehat O_1^{(3)} L =& -\fr{1}{2} \[1+ (-1)^m\] \[ \fr{n}{2} + \sum_{k<l} (-1)^{s_{kl}} \] (g^{uv})^{n-1} \mathcal{T}_{uv\cd uv} \\
    =& -\fr{1}{2} \[ \fr{n}{2} + \sum_{k<l} (-1)^{s_{kl}} \] g_{uv} L_1^{(1)}.
\end{aligned}
\end{equation}

Finally, consider $\widehat O_1^{(4)} L$ where
\be
\widehat O_1^{(4)} = -\fr{1}{2} \wtd R_{uv} \fr{\pa^2}{\pa \wtd R_{vv}\pa g^{vv}}.
\ee
This can be obtained from $\widehat O_1^{(3)} L$ by exchanging $u$ with $v$. This leads to
\be\label{eq:o4l}
\widehat O_1^{(4)} L = (-1)^m \widehat O_1^{(3)} L = \widehat O_1^{(3)} L.
\ee

Combining all four pieces of $\widehat O_1$, we find
\be
\begin{aligned}
    \widehat{O}_1 L =& \(\widehat O_1^{(1)}+\widehat O_1^{(2)}+\widehat O_1^{(3)}+\widehat O_1^{(4)}\) L \\
    =& \(\fr{n}{2} + \sum_{k<l} (-1)^{s_{kl}} - 2 \fr{1}{2} \[ \fr{n}{2} + \sum_{k<l} (-1)^{s_{kl}} \]\) g_{uv} L_1^{(1)} =0,
\end{aligned}
\ee
thus establishing the first relation.

\paragraph{Second relation}
We now prove the second relation
\be
\widehat O_2 L = 0
\ee
on the background. The calculation is similar to that of the first relation.

First, consider $\widehat O_2^{(1)} L$ where
\be
\widehat O_2^{(1)} = \wtd R_{uv} \fr{\pa}{\pa \wtd R_{uv}}.
\ee
On the background, we have
\be
\widehat O_2^{(1)} L = \widehat O_2^{(1)} L_2^{(1)}
\ee
where $L_2^{(1)}$ is equal to $L_1^{(1)}$ in \er{eq:L11}.  This gives
\be
\widehat O_2^{(1)} L = \fr{1}{2} (n-m) L_2^{(1)},
\ee
where $n-m$ is the number of $\wtd R$-contractions in the loop and the factor of $1/2$ comes from the symmetry of $\wtd R_{ab}$.

Second, consider $\widehat O_2^{(2)} L$ where
\be
\widehat O_2^{(2)} = \wtd R_{uv}^2 \fr{\pa^2}{\pa \wtd R_{uu} \pa \wtd R_{vv}}.
\ee
On the background, we have
\be
\widehat O_2^{(2)} L = \widehat O_2^{(2)} L_2^{(2)}
\ee
where $L_2^{(2)}$ is the sum of all terms in \er{eq:sing_loop} where the loop $(a_1b_1a_2b_2\cd a_nb_n)$ is alternating except for two defects at $\wtd R_{b_ka_{k+1}}=\wtd R_{uu}$ and $\wtd R_{b_l a_{l+1}}=\wtd R_{vv}$, for any $k \neq l$.  If $k < l$, compared to the alternating loop $(uvuv\cd uv)$ we are exchanging $u$ and $v$ in all $R$- and $\wtd R$-contractions between $g^{a_{k+1}b_{k+1}}$ and $g^{a_lb_l}$. Such a loop contributes
\be \label{eq:l3c2}
\fr{(-1)^{s_{k+1,l}}+(-1)^{s_{kl}}}{2} \fr{(-1)^{s_{k+1,l}}+(-1)^{s_{k+1,l+1}}}{2} (-1)^{s_{k+1,l}} (g^{uv})^{n} \fr{\wtd R_{uu}}{\wtd R_{uv}} \fr{\wtd R_{vv}}{\wtd R_{uv}} \mathcal{T}_{uv\cd uv}
\ee
to $L_2^{(2)}$. As with \er{eq:l3c}, this expression applies to any $l$ satisfying $k < l$, regardless of whether the $b_k,a_{k+1}$ and $b_l,a_{l+1}$ indices are contracted to some $\wtd R_{b_ka_{k+1}}$ and $\wtd R_{b_l a_{l+1}}$.

If $k>l$, compared to the alternating loop $(uvuv\cd uv)$ we are exchanging $u$ and $v$ in all $R$- and $\wtd R$-contractions \textit{not} between $g^{a_kb_k}$ and $g^{a_{l+1}b_{l+1}}$. Such a loop contributes
\be \label{eq:l3c2_2}
\fr{(-1)^{m-s_{l+1,k}}+(-1)^{m-s_{lk}}}{2} \fr{(-1)^{m-s_{l+1,k}}+(-1)^{m-s_{l+1,k+1}}}{2} (-1)^{m-s_{l+1,k}} (g^{uv})^{n} \fr{\wtd R_{uu}}{\wtd R_{uv}} \fr{\wtd R_{vv}}{\wtd R_{uv}} \mathcal{T}_{uv\cd uv}
\ee
to $L_2^{(3)}$.

Using the sum relations that we show in Appendix~\ref{app:details}, the prefactors in \er{eq:l3c2} and \er{eq:l3c2_2} after summing over $k,l$ simplify to
\begin{subequations}
\be \label{eq:l2sum1}
\sum_{k<l} \fr{(-1)^{s_{k+1,l}}+(-1)^{s_{kl}}}{2} \fr{(-1)^{s_{k+1,l}}+(-1)^{s_{k+1,l+1}}}{2} (-1)^{s_{k+1,l}} = \fr{m}{2} + \sum_{k<l} (-1)^{s_{kl}}
\ee
and
\begin{multline}
    \label{eq:l2sum2}
    \sum_{k>l} \fr{(-1)^{m-s_{l+1,k}}+(-1)^{m-s_{lk}}}{2} \fr{(-1)^{m-s_{l+1,k}}+(-1)^{m-s_{l+1,k+1}}}{2} (-1)^{m-s_{l+1,k}} \\
    = (-1)^m \[\fr{m}{2} + \sum_{k<l} (-1)^{s_{kl}}\].
\end{multline} 
\end{subequations}
Combining the two cases, we find
\ba
    \widehat O_2^{(2)} L &= \[1+ (-1)^m\] \[ \fr{m}{2} + \sum_{k<l} (-1)^{s_{kl}} \] (g^{uv})^{n} \mathcal{T}_{uv\cd uv} = \[ \fr{m}{2} + \sum_{k<l} (-1)^{s_{kl}} \] L_2^{(1)}.
\ea

Finally, consider $O_2^{(3)} L$ and $O_2^{(4)} L$ where
\be
\widehat O_2^{(3)} \equiv -\fr{1}{2} g^{uv} \wtd R_{uv} \fr{\pa^2}{\pa \wtd R_{uu}\pa g^{uu}},\qqu
\widehat O_2^{(4)} \equiv -\fr{1}{2} g^{uv} \wtd R_{uv} \fr{\pa^2}{\pa \wtd R_{vv}\pa g^{vv}}.
\ee
They were worked out in \er{eq:o3l} and \er{eq:o4l}, respectively. We therefore simply quote the results here:
\be
\widehat O_2^{(3)} L = \widehat O_2^{(4)} L = -\fr{1}{2} \[ \fr{n}{2} + \sum_{k<l} (-1)^{s_{kl}} \] L_2^{(1)}.
\ee
Combining all four pieces of $\widehat O_2$, we find
\be
\begin{aligned}
\widehat O_2 L =& \( \widehat O_2^{(1)} + \widehat O_2^{(2)} + \widehat O_2^{(3)} + \widehat O_2^{(4)} \) L \\
=& \(\fr{1}{2} (n-m) + \[ \sum_{k<l} (-1)^{s_{kl}} + \fr{m}{2} \] - \[ \fr{n}{2} + \sum_{k<l} (-1)^{s_{kl}} \] \) L_2^{(1)} = 0.
\end{aligned}
\ee \\

We have therefore proven that the two functions $f_{\text{SW}}$ and $f_{\text{EE}}$ are the same for any $f(\text{Riemann})$ theory, as claimed. This immediately implies our main result $v_B = \wtd{v}_B$ via \eqref{eq:vB} and \eqref{eq:vBtd}.

\section{Discussion}
\label{sec:discussion}
In this paper, we have shown that the butterfly velocity can be calculated using two distinct methods in holography: the shockwave method or the entanglement wedge method. We proved that the two methods give the same result for any $f(\text{Riemann})$ theory by direct computation. To find the butterfly velocity, we have solved the metric perturbation in the shockwave calculation and the near-horizon shape of extremal surfaces in the entanglement wedge calculation. In both methods, we have also taken a large-radius expansion in the transverse directions. After finding general expressions using both methods, their matching was not immediate. Nevertheless, exploiting the symmetry of the background solution on the horizon, we have shown that the difference indeed vanishes. 

While our calculations show explicitly that the two methods are equivalent for a large class of theories, a deeper and more intuitive understanding of the equivalence remains an interesting open question. In particular, the holographic entanglement entropy formula was derived by evaluating the gravitational action on a Euclidean conical geometry and varying it with respect to the conical angle~\cite{Lewkowycz:2013nqa,Dong:2013qoa,Camps:2013zua,Dong:2017xht,Dong:2019piw}, whereas the shockwave equation is derived in a Lorentzian spacetime with no conical defects. Furthermore, the shockwave profile \eqref{eq:h_ansatz} is exponentially decreasing in $r$, but the RT profile \eqref{eq:S_ansatz} is exponentially increasing in $r$. All these distinctions make the two methods appear very different, and finding a more direct way to connect them will likely shed light on the relationship between holographic entanglement and gravitational dynamics in general.\\

We now describe some potential future directions:

\paragraph{More general gravitational theories:} It would be interesting to see if the equivalence holds beyond $f($Riemann$)$ theories. To that end, we have worked out an example whose Lagrangian depends explicitly on the covariant derivative and found that the two methods continue to agree. More precisely, the Lagrangian contains
\be
    \mathcal{L} \supset \nabla_\mu R \nabla^\mu R,
\ee
and we find that its contributions to $f_{\text{SW}}(\mu)$ and $f_{\text{EE}}(\m)$ are equal (in $d=3$) and given by
\begin{multline}
    \frac{72 B_2 A_1}{A^4_0} +\frac{8 B_1 A_2}{A^4_0}-\frac{16 B_1 A^2_1}{A^5_0}-\frac{26 B^3_1}{A^3_0}-\frac{24 A_3}{A^4_0}
    -\frac{108 A^3_1}{A^6_0}+\frac{64 B_2 B_1}{A^3_0}-\frac{12 B^2_1 A_1}{A^4_0}\\+\frac{120 A_2 A_1}{A^5_0}-\frac{48 B_3}{A^3_0}-\bigg( \frac{8 B^2_1}{A^2_0}+\frac{4 A^2_1}{A^4_0}+\frac{12 B_1 A_1}{A^3_0} \bigg) \mu^2+\bigg(\frac{4 B_1}{A_0} +\frac{2 A_1}{A^2_0} \bigg) \mu^4.
\end{multline}
The holographic entanglement entropy functional for this theory can be found in \cite{Miao:2014nxa}. This example suggests that the two methods continue to agree in higher-derivative theories beyond $f($Riemann$)$. It would be interesting to prove this generally, including cases where the gravitational theory is coupled to matter fields with general interactions.  It would also be interesting to understand this better in the context of string theory, perhaps building on the results of~\cite{Shenker:2014cwa, Chandrasekaran:2021tkb}.

\paragraph{Beyond the butterfly velocity:} It would also be interesting to see if other properties of the OTOC related to shockwave quantities besides the butterfly velocity can be connected to properties of the entanglement wedge, further strengthening the link between gravity and entanglement. 

\paragraph{Connections to the Wald entropy:} An interesting connection between gravitational shockwaves and the Wald entropy was found in \cite{Liu:2021kay}. It was shown that the shockwave and microscopic deformations of the Wald entropy were related by a thermodynamic relation on the horizon. Since our main result establishes a connection between shockwaves and the generalized entropy \eqref{eq:S_EE}, it would be worth investigating to what extent their result can be related to ours.

\paragraph{Constraints on higher-derivative couplings:} As a potential application of our results, one could try to understand the constraints on higher-derivative couplings from the perspective of quantum chaos. To avoid issues related to unitarity and causality at finite couplings, we have treated the higher-derivative interactions perturbatively. The signs of these couplings appear constrained by the butterfly velocity. For example, in $d=2$ the butterfly velocity equals the speed of light in Einstein gravity. Therefore, requiring it be subluminal with higher-derivative corrections imposes constraints on the signs of the couplings. Given our expressions for a large class of higher-derivative theories, it would be interesting to see if requiring the butterfly velocity be subluminal can provide further constraints.

\paragraph{Relation to pole-skipping:} Throughout the paper we have focused on two methods of calculating the butterfly velocity --- the shockwave method and the entanglement wedge method. However, it has been suggested that the butterfly velocity (and more generally the OTOC) is also related to the phenomenon of pole-skipping \cite{Grozdanov:2017ajz, Blake:2017ris, Blake:2018leo}. In the gravitational context, this is related to the appearance of special points in Fourier space of the Einstein equations near the horizon, from which the Lyapunov exponent and butterfly velocity can be extracted. Although both the pole-skipping calculation and the shockwave method involve finding solutions to certain metric perturbations, the exact details are different. It would be interesting to explore their connections further.

\paragraph{Asymptotically flat spacetimes:}
Finally, both methods we discussed rely only on the near-horizon geometry and are therefore potentially generalizable beyond AdS spacetimes, such as asymptotically flat spacetimes, perhaps along the lines of \cite{Pasterski:2022lsl}.

\acknowledgments
We thank Gary Horowitz, Don Marolf, Mark Mezei, Jie-qiang Wu, and Ying Zhao for interesting discussions. X.D. and W.W.W. were supported in part by the Air Force Office of Scientific Research under award number FA9550-19-1-0360 and by funds from the University of California.  D.W. was supported by NSF grant PHY2107939.  C-H.W. was supported in part by the National Science Foundation under Grant No. PHY-1820908 and the Ministry of Education, Taiwan.

\begin{appendix}

\section{Exact linearity of the shockwave equation of motion}
\label{app:lin}
In this appendix, we show that there are no non-linear contributions, i.e., $\mathcal{O}(h^2)$, to the equation of motion from the shockwave perturbation in any higher-derivative theory of gravity including, but not limited to, $f($Riemann$)$. As an aside, we will also show that the only component of the equations of motion perturbed by the shockwave is $E^v_u$.

To achieve this, it will be useful to define a notion of \textit{chirality}.  Consider a (not necessarily covariant) tensor of the form $X^{a_1\cdots a_\ell}_{b_1\cdots b_n}$ built out of $g_{\mu\nu}$, $g^{\mu\nu}$ and $\partial_\mu$, where indices $a_1,\cd,a_\ell$, $b_1,\cd,b_n$ can be either $u$ or $v$, and we have suppressed $i$-type indices on $X$.  We define the chirality of any of its components as
\be
\chi = \#(v \text{ superscripts}) - \#(v \text{ subscripts}) - \#(u \text{ superscripts}) + \#(u \text{ subscripts}).
\ee
We refer to any tensor component with $\chi = 0$ as being non-chiral, and otherwise as being chiral. For example, the components $g_{uu}$ and $E^v_u$ are chiral since both have $\chi =2$, while $R_{uivj}$ is non-chiral since it has $\chi = 0$.

For all higher-derivative gravity theories, the equations of motion involve the metric, the Riemann tensor, and covariant derivatives. We can rewrite them using only the metric, the inverse metric, and partial derivatives. The only metric component that contains $h\d(u)$ is $g_{uu} = -2Ah\d(u)$; similarly, the only inverse metric component having $h\d(u)$ is $g^{vv} = 2A^{-1}h\d(u)$. A general term in $E^v_u$ therefore takes the form
\be\la{eq:eterm}
E^v_u \supset (\partial_v)^{N} X, \qqu
X = X_0 (\partial^{n_1}_u g_{uu}) (\partial^{n_2}_u g_{uu}) \cdots (\partial^{n_k}_u g_{uu}) (g^{vv})^m,
\ee
where we have collected all $v$-derivatives into the beginning of the expression (so that they are understood to act on particular parts of $X$ but not necessarily on $X$ as a whole), and collected everything that does not involve $g_{uu}$, its $u$-derivatives, or $g^{vv}$ into $X_0$. As $g^{uu}=g_{vv}=0$, $X_0$ is a product of $g_{uv}$, $\partial_u^\# g_{uv}$, and $g^{uv}$. Let $\c_0$ be the chirality of $X_0$; it is equal to the total number of $u$-derivatives, and thus always non-negative.  As $g_{uv}$ is a function of $uv$ only, each $\pa_u$ acting on $g_{uv}$ produces a factor of $v$, and we find
\be\la{eq:x0}
X_0 = v^{\c_0} f_0(uv)
\ee
where $f_0(uv)$ is some function of $uv$. Since $E^v_u$ has chirality 2, we need
\be\la{eq:cmatch}
\qqu N = 2m+2k-2+\sum_{i=1}^k n_i + \chi_0
\ee
for the chirality of the term in \er{eq:eterm} to agree.

Since $v$ appears only in the combination $uv$ in all metric functions, each $\partial_v$ in \er{eq:eterm} produces a factor of $u$ unless it acts on an explicit factor of $v$ produced by $\pa_u$.  In general, the $\pa_u$ acting on $g_{uu}=-2A(uv) h\d(u)$ in \er{eq:eterm} can act either on $A(uv)$ or the $\d$-function.

Let us first consider the simplest case where all $\pa_u$ shown in \er{eq:eterm} act on the $\d$-function.  In this case, using \er{eq:x0} we find
\be
X = v^{\c_0} f_1(uv) \d^{(n_1)}(u) \cdots \d^{(n_k)}(u) \(\d(u)\)^m,
\ee
where $f_1(uv)$ is some function of $uv$.  Therefore, the term \er{eq:eterm} in $E^v_u$ behave at most as
\be\la{eq:dist}
\wtd X \equiv (\partial_v)^N X \sim u^{N-\c_0} \d^{(n_1)}(u) \cdots \d^{(n_k)}(u) \(\d(u)\)^m,
\ee
keeping only the leading dependence on $u$.  Here we have acted as many $\partial_v$ as possible on $v^{\c_0}$; if not, we would get subleading contributions that are suppressed by additional powers of $u$. We will show momentarily that the leading contribution \er{eq:dist}, understood as a distribution, vanishes under the condition \er{eq:cmatch} unless it is actually $\d(u)$ or $u^n \d^{(n)}(u)$ for some $n$.  Thus any subleading contribution suppressed by additional powers of $u$ would always vanish as a distribution.

Now consider the more general case where not all $\pa_u$ shown in \er{eq:eterm} act on the $\d$-function.  Every $\pa_u$ that does not act on the $\d$-function must act on $A(uv)$ and produce an additional factor of $v$ (for one more $\pa_v$ to act on) --- thus the net effect on the term $\wtd X$ in \er{eq:dist} is to decrease one of the $n_i$ by $1$ and effectively increase $\c_0$ by $1$. This preserves the condition \er{eq:cmatch}, so it does not change our argument below.

We now show that the distribution \er{eq:dist} vanishes under the condition \er{eq:cmatch} unless it is actually $\d(u)$ or $u^n \d^{(n)}(u)$ for some $n$.  To see this, we regularize the $\d$-functions in \er{eq:dist} as narrow Gaussian functions:
\be
\d(u) \to \fr{\#}{\epsilon} e^{-u^2/\epsilon^2},
\ee
and integrate it against a test function $f(u)$:
\be
I \equiv \int_{-\infty}^\infty du \, \wtd X (u) f(u).
\ee
We find
\be
\begin{aligned}
I &\sim \int_{-\infty}^\infty du \, f(u) u^{N-\c_0} \(\fr{u}{\epsilon^2}\)^{\sum_{i=1}^k n_i} \fr{1}{\epsilon^{k+m}} e^{-\fr{(k+m)u^2}{\epsilon^2}} \\
&\sim \epsilon f(0) \epsilon^{2m+2k-2+\sum_{i=1}^k n_i} \epsilon^{-\sum_{i=1}^k n_i} \fr{1}{\epsilon^{k+m}} \\
&= f(0) \epsilon^{k+m-1},
\end{aligned}
\ee
where we have used \er{eq:cmatch} in going to the second line.  In the first line, we have written down a contribution to the regularized $\wtd X(u)$ where all $u$-derivatives act on the exponent of $e^{-u^2/\epsilon^2}$; every $u$-derivative that does not act on the exponent would remove a factor of $u^2/\e^2$ from the first line, but would not change the final result.

We now take the $\e\to0$ limit.  By construction, $k+m \geq 1$ since we are interested in corrections to the equations of motion due to the shockwave which have at least one factor of $\d(u)$ or its derivative. If $k+m >1$, then the integral $I$ vanishes as we send $\epsilon$ to zero. We are left with only two cases: either $k=0,m=1$ where $\wtd X \sim \d(u)$, or $k=1,m=0$ where $\wtd X \sim u^n \d^{(n)}(u)$ for some $n$. In either case, the term is a well-defined distribution and linear in $h$, concluding our proof for $E^v_u$.

Finally, consider other components of the equations of motion, e.g., $E^v_i$, $E^v_v$, etc.  They have $\chi \leq 1$, so we must have more powers of $\pa_v$ compared to \er{eq:cmatch}, and the corresponding distribution must have more powers of $u$ compared to \er{eq:dist}.  Thus the integral $I$ would go like at least $\mathcal{O}(\epsilon^{k+m})$, which vanishes in the $\e\to0$ limit as long as $k+m \geq 1$.  Therefore, other components of the equations of motion are not perturbed by the shockwave.

\section{Proof of sum relations} \label{app:details}
In this appendix, we prove the sum relations \er{eq:l2sum1} and \er{eq:l2sum2} used in the main proof. We will need the following identities
\begin{subequations}
\ba
\sum_{k=1}^n (-1)^{s_{kk}} &= \sum_{k=1}^n 1 = n, \label{eq:sum_id} \\
\sum_{k=1}^n (-1)^{s_{k,k+1}} &= \sum_{k=1}^n 1 + \sum_{k=1}^n \[(-1)^{s_{k,k+1}} - 1\] = n - 2 \sum_{k=1}^n \d_{s_{k,k+1},1} = n - 2 m. \label{eq:sum_id2}
\ea
\end{subequations}
In the sums below, the summation variables $k,l$ are always within the range $[1,n]$.

Beginning with \er{eq:l2sum1}, we prove it by writing
\ba\label{eq:sum1}
&\sum_{k<l} \[(-1)^{s_{k+1,l}}+(-1)^{s_{kl}}\] \[ (-1)^{s_{k+1,l}}+(-1)^{s_{k+1,l+1}} \] (-1)^{s_{k+1,l}} \nn\\
&= \sum_{k<l} \[1+(-1)^{s_{k,k+1}}\] \[ 1 + (-1)^{s_{l,l+1}} \] (-1)^{s_{k+1,l}} \nn\\
&= \sum_{k<l} (-1)^{s_{k+1,l}} + \sum_{k<l} (-1)^{s_{kl}} + \sum_{k<l} (-1)^{s_{k+1,l+1}} + \sum_{k<l} (-1)^{s_{k,l+1}} \nn\\
&= \(\,\sum_{k \leq l} (-1)^{s_{kl}} -\sum_{l=1}^n (-1)^{s_{1,l}} \) + \sum_{k<l} (-1)^{s_{kl}} + \sum_{k<l} (-1)^{s_{kl}} + \(\,\sum_{k \leq l} (-1)^{s_{k,l+1}} - \sum_{k=1}^n (-1)^{s_{k,k+1}}\) \nn\\
& \bmd
=\(\,\sum_{k < l} (-1)^{s_{kl}} +\sum_{k=1}^n (-1)^{s_{kk}} -\sum_{l=1}^n (-1)^{s_{1,l}} \) + \sum_{k<l} (-1)^{s_{kl}} + \sum_{k<l} (-1)^{s_{kl}} \\
+ \(\,\sum_{k < l} (-1)^{s_{kl}} +\sum_{k=1}^n (-1)^{s_{k,n+1}} - \sum_{k=1}^n (-1)^{s_{k,k+1}}\) 
\emd \nn\\
&= 4 \sum_{k < l} (-1)^{s_{kl}} + \sum_{k=1}^n \big[ (-1)^{s_{kk}} - (-1)^{s_{k,k+1}} \big] \nn\\
&= 4 \sum_{k < l} (-1)^{s_{kl}} + 2m,
\ea
where we have used $s_{1,k}=s_{k,1}=s_{k,n+1}$ in going to the second-to-last line, and used \er{eq:sum_id} and \er{eq:sum_id2} in going to the last line.

Similarly, we prove \er{eq:l2sum2} by writing
\ba
    &\sum_{k>l} \[(-1)^{m-s_{l+1,k}}+(-1)^{m-s_{lk}}\] \[(-1)^{m-s_{l+1,k}}+(-1)^{m-s_{l+1,k+1}}\] (-1)^{m-s_{l+1,k}}\nn \\
    &= (-1)^m \sum_{k>l} \[(-1)^{s_{l+1,k}}+(-1)^{s_{lk}}\] \[(-1)^{s_{l+1,k}}+(-1)^{s_{l+1,k+1}}\] (-1)^{s_{l+1,k}} \nn\\
    &= (-1)^m \[ 4 \sum_{k < l} (-1)^{s_{kl}} + 2m \]
\ea
where in going to the last line we have used the fact that the sum is the same as \er{eq:sum1} with $k \lra l$.

\end{appendix}

\bibliographystyle{JHEP}
\bibliography{bibliography}

\end{document}